\documentclass[fleqn,usenatbib,useAMS]{mnras}

\usepackage{graphicx}	
\usepackage{amsmath}	
\usepackage{amssymb}	
\usepackage{multicol}        
\usepackage{bm}		
\usepackage{pdflscape}	

\newcommand{\kms}{\,km\,s$^{-1}$} 
\usepackage[T1]{fontenc}
\usepackage{ae,aecompl}
\usepackage{mathptmx}

\title[]{SDSS-IV MaNGA: Stellar initial mass function variation inferred from Bayesian analysis of the integral field spectroscopy of early type galaxies}

\author[S. Zhou et al.]
{Shuang Zhou$^{1}$\thanks{Contact e-mail: \href{mailto:zhou-s13@mails.tsinghua.edu.cn}{zhou-s13@mails.tsinghua.edu.cn}},
H.J. Mo$^{1,2}$,
Cheng Li$^{1}$,
Zheng Zheng$^{3}$,
Niu Li$^{1}$,
Cheng Du$^{1}$,
Shude Mao$^{1,3,7}$,
\newauthor
Taniya Parikh$^{4}$,
Richard R. Lane$^{5,6}$,
Daniel Thomas$^{4}$
\\
$^{1}$Tsinghua Center of Astrophysics $\&$ Department of Physics, Tsinghua University, Beijing 100084, China\\
$^{2}$Department of Astronomy, University of Massachusetts Amherst, MA 01003, USA\\
$^{3}$National Astronomical Observatories, Chinese Academy of Sciences, A20 Datun Road, Chaoyang District, Beijing 100012\\
$^{4}$Institute of Cosmology $\&$  Gravitation, University of Portsmouth, Dennis Sciama Building, Portsmouth, PO1 3FX, UK\\
$^{5}$Instituto de Astrof\'{\i}sica, Pontificia Universidad Cat\'{o}lica de Chile, Av. Vicu\~na  Mackenna 4860, 782-0436 Macul, Santiago, Chile\\
$^{6}$Millennium Institute of Astrophysics, Av. Vicu\~na Mackenna 4860, 782-0436 Macul, Santiago, Chile\\
$^{7}$Jodrell Bank Centre for Astrophysics, Alan Turing Building, The University of Manchester, Manchester M13 9PL, UK\\
}
\date{Last updated XXX; in original form XXX}

\pubyear{2018}

\begin{document}
\label{firstpage}
\pagerange{\pageref{firstpage}--\pageref{lastpage}}
\maketitle

\begin{abstract}
We analyze the stellar initial mass functions
(IMF) of a large sample of early type galaxies (ETGs)
provided by MaNGA. The large number of IFU spectra of individual galaxies
provide high signal-to-noise composite spectra that are
essential for constraining IMF and to investigate possible
radial gradients of the IMF within individual galaxies.
The large sample of ETGs also make it possible to study how
the IMF shape depends on various properties of galaxies.
We adopt a novel approach to study IMF variations in ETGs,
use Bayesian inferences based on full spectrum fitting.
The Bayesian method provides a statistically rigorous way to
explore potential degeneracy in spectrum fitting, and to
distinguish different IMF models with Bayesian evidence.
We find that the IMF slope depends systematically on
galaxy velocity dispersion, in that galaxies of higher
velocity dispersion prefer a more bottom-heavy IMF, but the
dependence is almost entirely due to the change of
metallicity, $Z$, with velocity dispersion. The IMF shape also
depends on stellar age, $A$, but the dependence is
completely degenerate with that on metallicity through a
combination $AZ^{-1.42}$. Using independent age and metallicity
estimates we find that the IMF variation is produced by
metallicity instead of age. The IMF near the centers of massive
ETGs appears more bottom-heavy than that in the outer parts, while
a weak opposite trend is seen for low-mass ETGs. Uncertainties
produced by star formation history, dust extinction, $\alpha$-element
abundance enhancement and noise in the spectra are tested.
\end{abstract}



%
\section{Introduction}

Stars are the building blocks of galaxies. An accurate description of the
distribution and properties of the stellar components of galaxies
plays a key role in understanding galaxy structure, formation and evolution.
Because of this, the stellar initial mass function (IMF), which describes
the mass distribution of stars at birth,  has been the subject of
numerous investigations. More than half a century ago,
\cite{Salpeter1955} proposed a model of IMF, which is a single power
law, $\propto m^{-2.35}$, over the entire mass range.
However, subsequent studies of resolved stellar populations in our own Galaxy reveal
a flattening of the IMF at the low-mass end, $m<0.5\,{\rm M}_\odot$,
and the IMF is found to be better described by the forms given by \citet{Kroupa2001}
and \citet{Chabrier2003}. These forms have been widely used in modeling
stellar populations of the Milky Way galaxy as well as of external
galaxies.

 One of the most important questions in the field of galaxy evolution,
which remains a hot issue of debate up to now, is whether the form of the IMF is
universal or varies from galaxy to galaxy. With the advent of
detailed spectroscopic and photometric observations of galaxies,
as well as more accurate stellar population models, it is now possible
to address the question of IMF variation with observational data.
There are now increasing amounts of evidence showing that the IMF
of early-type galaxies (ETGs) deviates significantly from the
forms suitable for the Milky Way.
Investigations based on detailed dynamical modeling of galaxies
\citep{Thomas2011,Dutton2012,Cappellari2012,Cappellari2013,Lyubenova2016,lhy2017}, and gravitational
lensing effects of ETGs \citep{Treu2010,Posacki2015,Newman2017},
have indicated that the masses of ETGs appear to be
larger than that implied by their luminosities together with
a Milky Way type IMF. The observational data thus seem to prefer
an IMF that contains an excessive amount of objects with high mass-to-light ratios.
However, the lensing and
dynamical results only provide constraints on the overall
mass-to-light ratios of individual galaxies, which can be increased
either by an excess of low-mass stars or by stellar remnants from
giant stars \citep{Cappellari2013}. Moreover, some nearby, strongly lensed, massive galaxies,
which have exceptionally accurate mass determinations, are found
to be consistent with a Kroupa-like IMF
\citep{Smith2013,Smith2015,Collier2018}, suggesting that
the conclusion of a bottom-heavy IMF for ETGs from such
studies is still uncertain.

Since stars of different masses have different spectra, both in
continuum shapes and in absorption features, one can, in principle,
obtain information about the IMF of galaxies by studying their spectra.
Many of the earlier investigations along this line are based on
IMF-sensitive absorption lines or features in galaxy spectra.
 For example, \cite{Cenarro2003} suggested a significant anti-correlation
between Ca II triplet index and central velocity dispersion using
a sample of 35 early-type galaxies, which indicates an excess of low-mass stars
in massive galaxies. Such an enhancement of dwarf-sensitive absorption
features in massive ETGs was confirmed by \cite{vanDokkum2010,vanDokkum2011}
using a sample of eight massive ETGs in the Virgo and Coma clusters.
To model these features, \citet{Conroya2012}
developed new stellar population models and techniques to fit the local spectral
shapes near the absorption features, and applied them  to  a set of 34 ETGs
from the SAURON survey \citep{vanDokkum2012,Conroyb2012}. These techniques have since been widely adopted
to analyze IMF variations \citep[e.g.][]{Villaume2017,vanDokkum2017, Vaughan2018}.
In addition to modeling the exact shapes of absorption features, measurements of
line strengths are also used to quantify the absorption features,
and to study the variations of IMF \citep{Spiniello2012,Spiniello2014,Spiniello2015,
Ferreras2013,Barbera2013,Barbera2015,Navarrob2015}. All these investigations demonstrated that the IMFs of massive ETGs are
significantly different from that of the Milky Way, and that there
are systematic variations of the IMF with galaxy properties,
such as velocity dispersion and metallicity.

Efforts have also been made to study possible spatial variations of
the IMF shape within individual galaxies. However, it is time-consuming and
expensive to obtain spatially resolved spectra that reach the quality
needed for IMF analysis, only a handful of investigations have been
carried out so far and no consensus has yet been reached.
Some investigations show that the IMF within massive ETGs changes
systematically from strongly bottom-heavy in the center
to a more mildly bottom-heavy in the outer part
\citep[e.g.][]{Navarrob2015,Barbera2017,vanDokkum2017,Vaughan2018b,Sarzi2018,Taniya2018}. However,
other investigations question the results, suggesting that the
radial variation in the stellar population is generated
by abundance gradients of individual elements,  rather
than by a change in the IMF
\citep[e.g.][]{Zieleniewski2015,Zieleniewski2017,McConnell2016,Alton2017,Vaughan2018}.

In addition to studies focusing on the low-mass end of the IMF
in ETGs, there are also attempts to study the IMF in the high mass end.
For example, \cite{Zhang2007} used emission features of Wolf-Rayet stars to
study the population of massive stars, and found that the IMF is correlated
with the gas phase metallicity of galaxies. \cite{Meurer2009} measured the
flux ratio ${\rm F_{H\alpha}/F_{FUV}}$ in a sample of HI-selected galaxies
and reported the finding of a systematic variation of the IMF with galaxy
surface brightness. \cite{Gunawardhana2011} observed strong dependence of
the the IMF on star formation in a sample of low-to-moderate star-forming
galaxies. If these star-forming galaxies have properties similar to
the progenitors of ETGs, the IMF variations observed might be related to
those observed in ETGs today.

Theoretical ideas have also been proposed to understand
the origin of the observed IMF variations, although no census has
been reached. \cite{Chabrier2014} studied the Jeans mass
in a turbulent medium and found that the IMFs in very dense
and turbulent environments tend to be bottom-heavy.
\cite{Jerabkova2018} calculated the galaxy-wide IMF (gwIMF) from the
IGIMF-theory (IMF based on resolved star clusters) and found that
the the shape of gwIMF depends on both metallicity and
star formation rate. There are also attempts to reproduce
simultaneously the observed IMF variation, mass-metalicity relation
and abundance patterns, either by assuming a time-dependent form of
the IMF \citep{ Weidner2013, Ferreras2015} or using a chemical
evolution model \citep{ DeMasi2018}. The influences of such variations
on the interpretation of the observed galaxy populations have also
been investigated using analytical calculations \citep[e.g.][]{Clauwens2016}
and cosmological simulations
\citep[e.g.][]{ Gutcke2019, Barber2018, Barber2019a, Barber2019b}.

In this paper we intend to carry out a systematic analysis of
the IMF of ETGs using a large sample constructed from the Mapping
Nearby Galaxies at Apache Point Observatory (MaNGA; \citealt{Bundy2015}).
The large number of IFU spectra of individual galaxies provided by MaNGA
not only allow us to obtain high signal-to-noise composite spectra for
individual galaxies, which is essential for constraining IMF, but also allow
us to study spatial variations (e.g. radial gradient) of the IMF
within individual galaxies. The large sample of ETGs also make it
possible to study how the IMF shape depends on various properties
of galaxies. Indeed, MaNGA data has already been used to study IMF variations
through dynamical constraints \citep{lhy2017} and absorption
line indexes  \citep{Taniya2018}. In contrast to those earlier investigations, our analysis
is based on our newly developed stellar population synthesis (SPS) code,
Bayesian Inference for Galaxy Spectra (BIGS), which fits the full
composite spectrum of a galaxy and to constrain its IMF shape
along with other properties of its stellar population, such as
age and metallicity. The Bayesian approach also provides a
statistically rigorous way to explore potential degeneracy in
model parameters inferred from the spectrum fitting, and to
distinguish different IMF models through Bayesian evidence.

The paper is organized as follows. In \S\ref{data} we present our
data reduction process, including sample selection and spectral
stacking procedure. \S\ref{analysis} provides a brief introduction
to the SPS model and Bayesian approach used to fit galaxy spectra.
Our results are presented in \S\ref{result}, followed by some
discussions about potential uncertainties in them. Comparisons
with earlier results are made in \S\ref{sec_comparisons}, and
our main conclusions are summarized in \S\ref{summary}.

%
\section{DATA}
\label{data}

\subsection{The MaNGA survey}

MaNGA is one of the three core programs in the fourth-generation Sloan
Digital Sky Survey (SDSS-IV, \citealt{Blanton2017}). As a new IFU spectroscopic
survey, MaNGA aims to observe the internal kinematic structure and composition of gas
and stars in an unprecedented sample of about 10,000 nearby galaxies. Targets are chosen
from the NASA Sloan Atlas catalogue
\footnote{\label{foot:nsa}\url{ http://www.nsatlas.org/}}
(NSA, \citealt{Blanton2005}) to cover the stellar mass range
$5\times10^8 M_{\odot}h^{-2} \leq M_*\leq 3 \times 10^{11} M_{\odot}h^{-2}$, with median
redshift $z\sim0.03$ and a roughly flat number density distribution \citep{wake2017}.
The sample consists of three subsamples: the Primary sample, Secondary sample
and Colour-Enhanced sample. The Primary and Secondary samples are selected
to be flat in the distribution of the $K$-corrected $i$-band magnitude,
and have a spatial coverage of at least $1.5 R_e$ and $2.5 R_e$
($R_e$ being the effective radius of a galaxy) within the
IFUs, respectively. The Colour-Enhanced sample selects galaxies in
regions that are not well sampled by the Primary sample in the
NUV-$i$ color versus $M_i$ plane.

With the help of the two dual-channel BOSS spectrographs
\citep{smee2013} on the Sloan 2.5 m telescope \citep{Gunn2006}, MaNGA provides
simultaneous wavelength coverage
over $3600-10300$ {\AA}, with a spectral resolution $R\sim2000$ \citep{Drory2015},
reaching a typical $r$-band signal-to-noise ratio (SNR) of $4-8$
({\AA}$^{-1}$ per $2''$ fiber) at 23 AB mag arcsec$^{-2}$ in the outskirts of
MaNGA galaxies. We refer the reader to \cite{Law2015} for the observing strategy,
to \cite{Yana2016} for the details of the spectrophotometry calibration,
and to \cite{Yanb2016} for the initial performance of MaNGA.

\subsection{Sample selection}

\begin{figure}
\includegraphics[height=70mm]{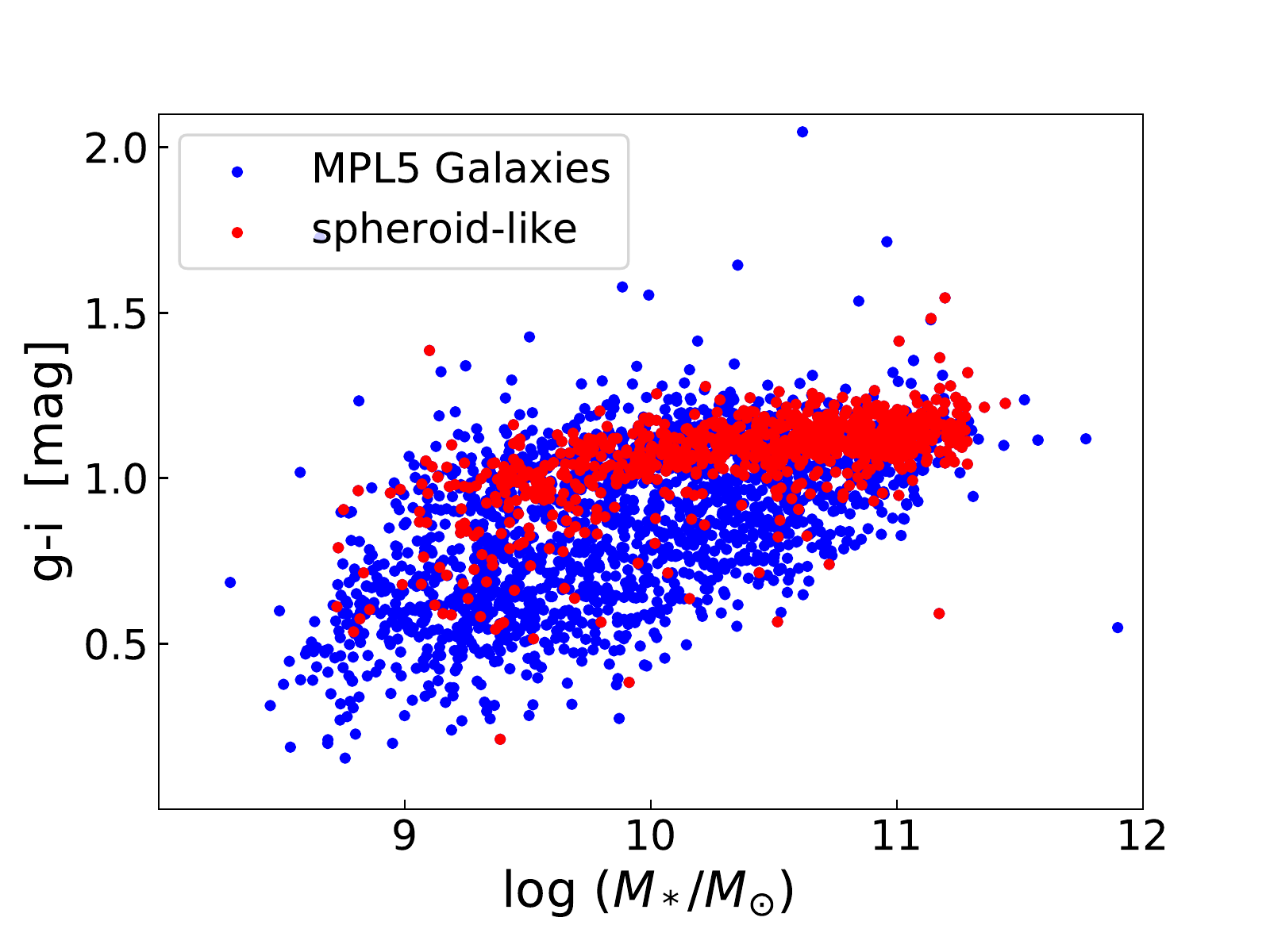}\\
\caption{Stellar mass-colour distribution of MPL5 galaxies.
Galaxies visually identified as `spheroid-like', which are used
in our analysis, are shown in red dots, while the other two classes,
disk-like and irregular, are shown by the blue dots.
The $g-i$ color and the total stellar mass $M_*$ are adopted from
NSA catalog.
}\label{color}
\end{figure}

In our analysis we use the MaNGA MPL5 data, which includes a total of 2778
galaxies from the first two years of MaNGA survey operations, and is released
in the SDSS fourteenth data release \citep[SDSS DR14][]{Abolfathi2018}.
We adopt the morphology classification from \cite{Wang2018}, in which the
morphological type of each galaxy is classified into three broad classes:
disk-like, spheroid-like and irregular, by visually inspecting its $r$-band
image in the SDSS. Our analysis focuses on the spheroid-like class. After excluding galaxies with apparent problems in data reduction
and/or spectral fitting, we obtain a sample of 894 galaxies.
The spheroid-like object defined here is a stricter
classification of ETG than in the widely-used Galaxy-Zoo classification
\citep{Willett2013}. Some ETGs in Galaxy-Zoo classification have
spiral-like features in their image, and thus are dropped from our
sample \citep{Wang2018}. We have also compared our sample with the ETGs
from a value-added catalogue of MaNGA galaxies presented
in \cite{Fischer2019}. Although the two samples have differences in
identifying S0 galaxies, they are very similar in their mass
and color distributions. Such differences are not expected to affect
our main results. Fig.\,\ref{color}
shows the distributions of the MPL5 galaxies and of our sample galaxies
in the $g-i$ colour versus $M_*$ plane, where both $M_*$ and $g-i$ are obtained
from NSA. The total stellar masses of galaxies provided by NSA are
derived from the fit to the SDSS five-band photometry with K-corrections
\citep{Blanton2007}, using the \citet{BC03} model and the \citet{Chabrier2003}
IMF.

\subsection{Data reduction}

The IFU spectra used in the paper are extracted using the MaNGA official data reduction
pipeline (DRP, \citealt{Law2016}). DRP produces sky-subtracted, spectrophotometrically
calibrated spectra, and combines individually dithered observations into three
dimensional data cubes. Relative flux calibration for the MaNGA data is better
than $5\%$ \citep{Yana2016}. Additional information adopted in our analysis comes
from the MaNGA Data Analysis Pipeline (DAP, Westfall et al., in preparation).
This pipeline uses publicly available code, pPXF
\citep{Cappellari2004,Cappellari2017}, to fit the stellar continuum and nebular
emission lines of each spectrum, returning the kinematics of both components.
We use their results for the stellar components, including the stellar
velocity and velocity dispersion maps, as well as the elliptical polar
radius of each spaxel from the galaxy center, normalized by the elliptical
Petrosian effective radius from the NSA.

\subsection{Stacking of spectra}
\label{stack}

In general, the flux variation caused by different IMFs is only at
a few percent level \citep[e.g.][]{Conroya2012} due to the
large mass-to-light ratio of dwarf stars \citep[e.g.][]{Maraston1998}.
Thus, to analyze the IMF variations generally requires spectral
signal to noise ratio (SNR) of at least 100 {\AA}$^{-1}$
\citep[e.g.][]{Ferreras2013}, which is often achieved only by stacking spectra
from a sample of galaxies, prohibiting such analyses for individual galaxies.
The original DRP spectra, with typical $r$-band SNR of $4-8$ {\AA}$^{-1}$ in
the outer parts of galaxies, are obviously not sufficient for
detailed analyses of IMF variations within individual galaxies.
However, with the help of thousands of spectra in each IFU plate,
one can obtain galaxy spectra with sufficiently high
SNR by combining the spaxels of individual galaxies.

In this paper, we study two kinds of stacked spectra for different purposes.
First, to study the global properties of individual galaxies and the
variations of these properties from galaxy to galaxy, we co-add all
the spaxels inside one effective radius ($R_e$, from NSA) of a galaxy into
a single spectrum. Since the SNR is expected to increase by a factor
of $N^{1/2}$, with $N$ being the number of spaxels, the stacked spectra of
individual MaNGA galaxies can reach the SNR needed
to study IMF effects. Unfortunately, spaxels from a single
galaxy are still insufficient to reach the required SNR for
a detailed investigation of the radial gradient of its IMF.
To study potential radial variations in the IMF, therefore,
requires a different kind of stacking. Here we first divide our sample
into four subsets according to galaxy stellar mass, with
$\log(M_*/M_{\odot})\in[8.0,10.0]$, $[10.0,10.5]$, $[10.5,11.0]$,
and $[11.0,12.0]$, respectively. In each mass bin, the spaxels within $1 R_e$
for individual galaxies are separated into 10 radial bins according to
their normalized radii of elliptical annuli, using a bin size of $0.1 R_e$.
These radial bins from different galaxies are then combined
accordingly to produce a final set of $4\times10$ spectra to be
used for the analyses of the radial dependence.

The binning procedure is as follows. Before performing the stacking,
we take advantage of the high quality masks produced by the DRP and DAP to
get rid of all the quantities that are problematic, such as low or no
fibre coverage, and foreground star contamination.
These masked individual pixels in the spectra from DRP, and the corresponding
outputs of stellar velocity, $v$, and velocity dispersion,
$\sigma_*$, from DAP are excluded. To do the stacking, we first convert the
wavelengths of each spectrum to the rest-frame, using the redshift $z$
of each galaxy as given by NSA and the stellar velocity $v$ from the DAP.
A cut at ${\rm SNR}=10$ is used to ensure the accuracy in the
measurements of kinematics. The flux, inverse-variance and spectral
resolution vectors from DRP are then interpolated to a common wavelength
grid, uniformly sampled in the logarithmic space. For spaxels from the same
galaxy, we calculate the direct mean of the flux at each wavelength point,
which corresponds to a light-weighted average flux. For the stacking of
radial bins of different galaxies, we first normalize the spectra in
the wavelength window $4500-5500$ {\AA}, which takes into account
the flux variations among galaxies, and then calculate the
inverse-variance weighted average of the flux at each wavelength
point to maximize the SNR. The error vectors are then generated from
the error propagation formula. For the velocity dispersion, we use the
quadratic mean of the velocity dispersions measured from individual
spectra as an initial estimate for the stack.

\begin{figure}
\includegraphics[height=70mm]{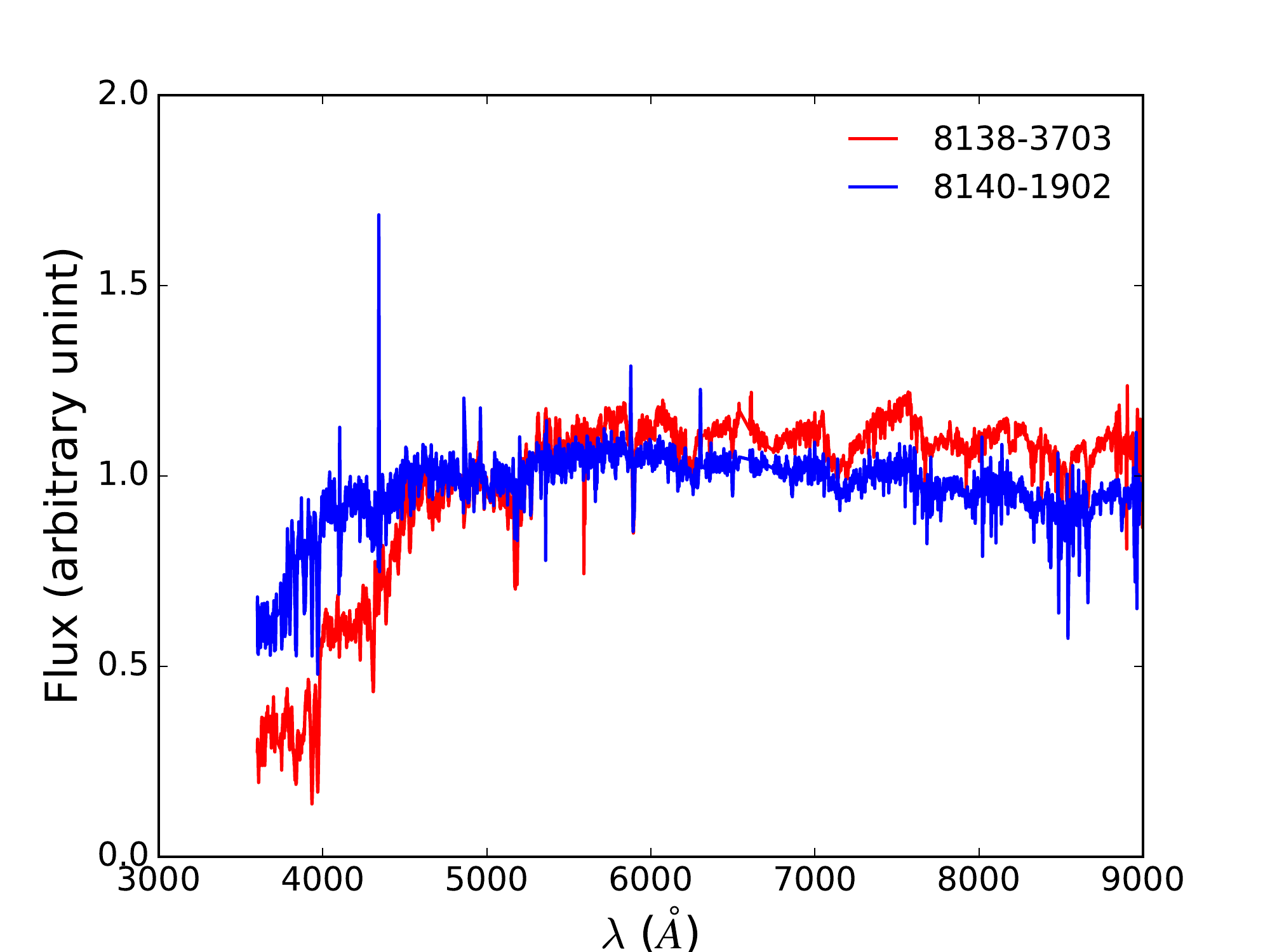}\\
\includegraphics[height=70mm]{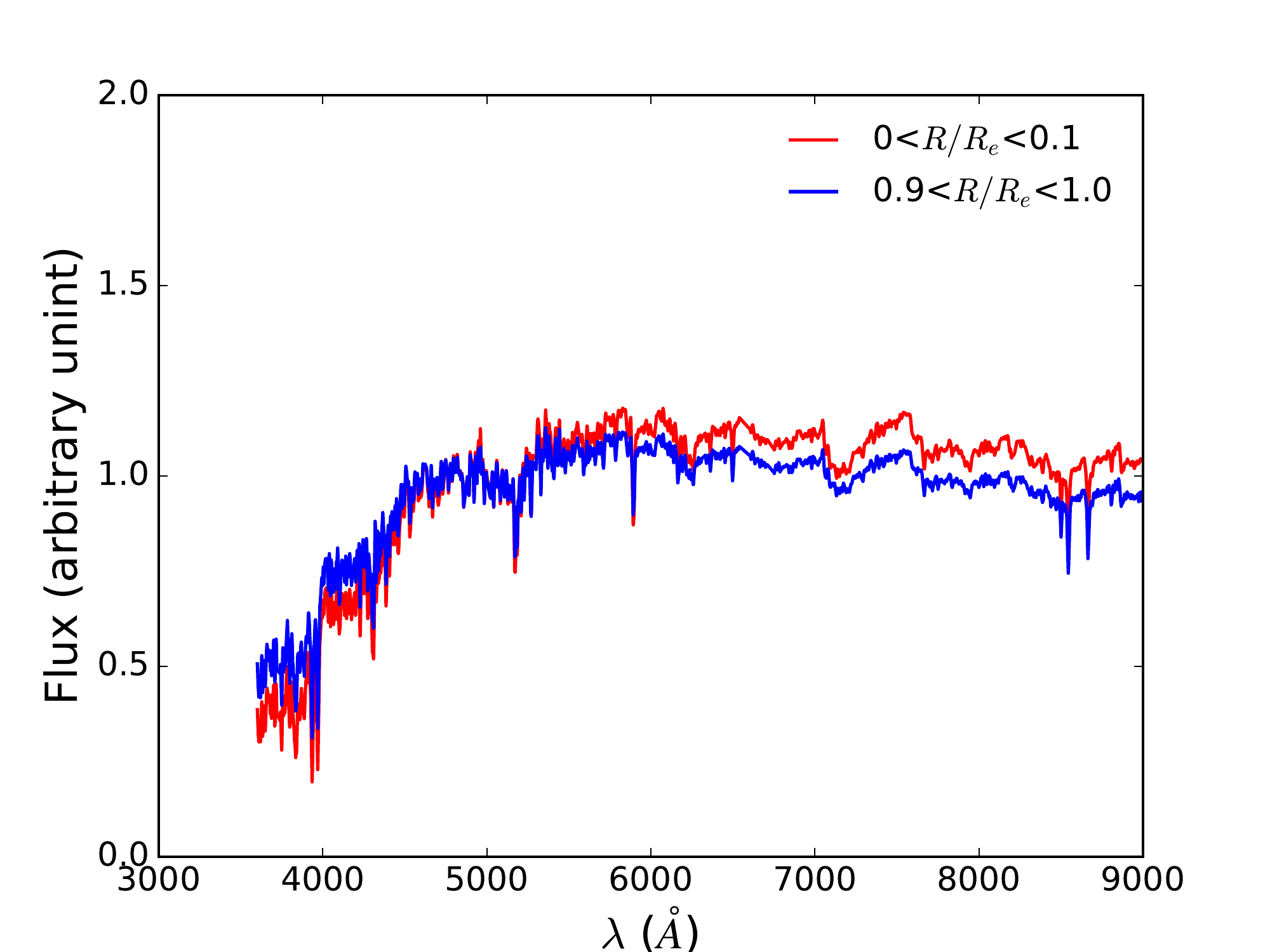}\\
\caption{Top panel: stacked spectra within $1 R_e$ for two ETGs with NSA central
velocity dispersion $\sigma_*$ equal to 230\kms (red) and 74 \kms (blue),
respectively. Bottom panel: radially binned spectra for the most massive sample
(NSA $M_*>10^{11} {\rm M}_{\odot}$) in the central (red) and outer (blue)
radial bins.}
\label{spectra}
\end{figure}

Examples of the spectra finally used in the paper are shown in
Fig \ref{spectra}. As one can see, the SNR of the
stacked spectra of individual galaxies within $1R_e$ varies from
galaxy to galaxy. The typical SNR of the stacked spectra
are around 200 pixel$^{-1}$, with wavelength dependence that
peaks at around 6000 {\AA} and drops towards both the
red and blue ends. High mass galaxies typically have old red spectra,
while lower mass galaxies have bluer spectra, with emission lines.
For the radially stacked spectra of a sample of galaxies, the central regions are
typically redder than the outer parts. The SNR is
significantly enhanced in the stacked spectra, and is
at least 500 pixel$^{-1}$ even for the outer bins.
It should be noted, however, that some sky line residuals, especially in the near-IR
wavelength range, can be seen in the spectra of individual galaxies,
which may potentially bias our IMF inferences. However, these residual
contaminants are greatly reduced in the stacked spectra,
apparently because they are averaged out when the spectra of
galaxies at different redshifts are corrected to the rest frame.
To make full use of this crucial benefit, we obtain the
average properties of a sample of galaxies from the stack
spectra, rather than from the mean of the quantities
derived from individual spectra.

%
\section{Analysis}
\label{analysis}

\begin{figure}
\includegraphics[height=70mm]{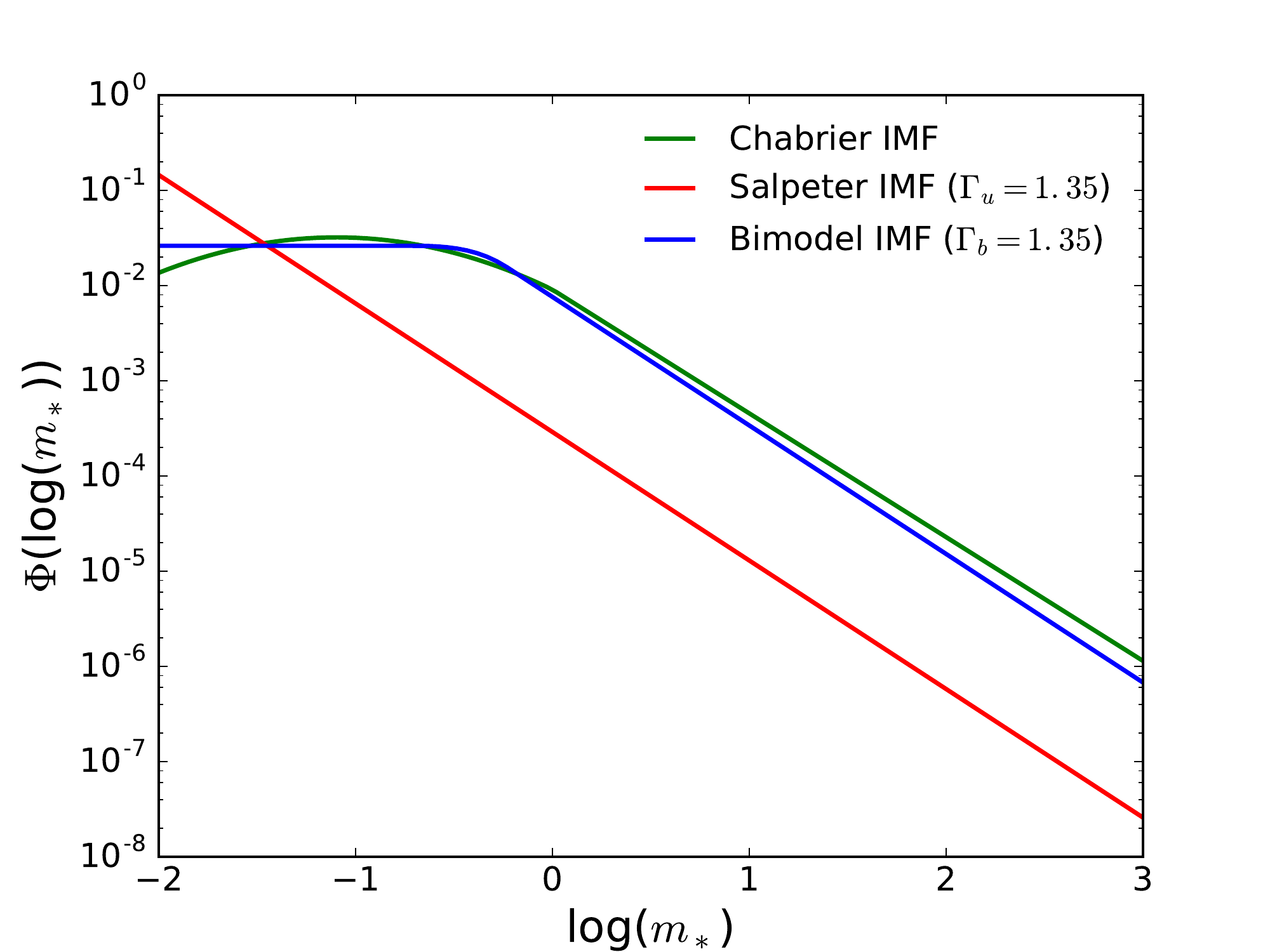}\\
\caption{Examples of IMF models used in our analysis, as labeled.}
\label{IMF}
\end{figure}

To infer properties of the IMF from galaxy spectra, one needs to compare stellar
population models with the observed spectra, either by using some
IMF-sensitive features or by fitting an entire spectrum. These two
approaches are complementary. In general, the full spectrum contains more
information, but it may be difficult to disentangle different factors
that can affect the spectrum of a galaxy. In contrast,
absorption features selected may be sensitive only to
specific stellar populations, which may miss important information
contained in the spectrum. In this paper, we adopt the approach based
on full spectrum fitting, using an approach based on
Bayesian statistics to infer the IMF and its co-variances
with other properties.

\subsection{The spectral synthesis model}

The stellar population synthesis approach has long been adopted
as the standard tool to interpret galaxy spectra \citep[see][for details]{Conroy2013}.
In this approach, the time evolution of stars of different masses from the zero-age
main sequence to their deaths is used to compute isochrones, which are the
age-dependent distribution of stars in the Hertzsprung-Russell (HR)
diagram. With the use of empirical and/or empirically-calibrated theoretical stellar
libraries to assign a spectrum to each point in the HR diagram,
the integrated light from a coeval set of stars (referred to as a simple
stellar population, SSP) can be obtained by summing up the spectra along an
isochrone, weighted by the number of stars at a given stellar mass, as
specified by the IMF of stars. Finally, for a composite stellar population
(CSP), such as a galaxy, the spectrum can be obtained through
the convolution of the time-dependent SSPs with the star formation
history (SFH) of the galaxy, together with a prescription for dust
attenuation. In what follows we describe briefly the SSP models we use.

\subsubsection{Models of \ SSP}
\label{SSP}

There are several popular SSP models, e.g.  BC03 \citep{BC03},
M05 \citep{Maraston2005}, CvD12 \citep{Conroya2012} and
E-MILES \citep{Vazdekis2016}, that are based on different stellar
templates and isochrones. Each of these models has its own merits and
shortcomings. Since we are interested in the inference of the IMF, whose
effects are the strongest in the near infrared (NIR) \citep{Ferreras2013},
we need to select SSP models that can provide consistent
predictions all the way to the NIR. This kind of SSP models are currently
rare and calibration may be a problem in many of them.
In this paper, we select a set of state-of-the-art
E-MILES\footnote{\label{foot:emiles}\url{http://miles.iac.es/}}
models as our SSP templates.

The E-MILES models are the latest version of the MILES models originally presented in \cite{Vazdekis2010}.
They are constructed using the MILES \citep{miles2006MNRAS},
CaT \citep{Cenarro2001} and Indo-U.S. \citep{Valdes2004} empirical
stellar libraries, extended  to the infrared with the use of the IRTF
stellar library \citep{Cushing2005,Rayner2009}.
Self-consistent E-MILES SSP spectra are computed to cover the wavelength range
from 1680.2 {\AA} to $5{\rm \mu m}$, with a moderately high spectral resolution.
In particular, these SSPs can reach a high resolution of 2.51 {\AA} (FWHM)
over the range from 3540 {\AA} to 8950 {\AA}, which is of special interest
in our analysis. Since the wavelength range of the original MaNGA data is
$3600-10300$ {\AA}, most of the spectra will fall within this range
after corrected to the restframe. We thus limit our analysis of the stacked
spectra only to this wavelength range.

The E-MILES SSPs are computed for several IMFs, including the widely used
Salpeter \citep{Salpeter1955}, Kroupa \citep{Kroupa2001} and
Chabrier \citep{Chabrier2003} IMFs, and two parameterized forms,
unimodel and bimodel \citep{Vazdekis1996}, both including
a lower and an upper mass-cutoff set at $0.1M_\odot$
and $100M_{\odot}$, respectively. The unimodel IMF is a simple power
law over the entire mass range, characterized by the power law slope,
$\Gamma_u$:
\begin{equation}
\Phi(m)\propto m^{-(\Gamma_u+1)}
~~~~(\text{unimodel})\,.
\end{equation}
Thus, the Salpeter IMF is a special case of this kind with
$\Gamma_u = 1.35$. In contrast, the bimodel IMF is the same as
the unimodel IMF at $M_*>0.6 M_{\odot}$, characterized by a similar
power law slope $\Gamma_b$, but turns to a flat
distribution at the lower mass end:
\begin{equation}
\Phi(m)\propto
 \left\{\begin{array}{lr}
   (m/0.6)^{-(\Gamma_b+1)} &\text{(for $m>0.6$)}\\
   p(m)                    &\text{(for $0.2\leq m\leq0.6$)}\\
   1                       &\text{(for $m<0.2$)}
   \end{array}
~~~(\text{bimodel})\,.
   \right.
\end{equation}
where $p(m)$ is a third degree spline between the low- and high-mass
ends of the IMF. The relatively low fraction of low-mass stars in this IMF is
in agreement with that inferred from the Milky Way. In fact, a bimodel IMF
with $\Gamma_b = 1.3$ is a good approximation to the Kroupa IMF.
Some examples of these IMFs are shown in Fig \ref{IMF}.
In what follows, we will compare the Milky-way like and bottom-heavy IMFs
in our model selection analysis, while using unimodel and bimodel to investigate
the shape of the IMF.

\subsubsection{Star formation history and dust extinction}

The star formation history (SFH) of a galaxy can in principle be
very complex, and it is difficult to come up with a universal model
that can describe the SFHs of every individual galaxies.
Furthermore, the SFH may also vary across a galaxy,
which makes it even more difficult to model the spectra
in an IFU survey such as  MaNGA. Currently there are two ways
to model the SFH. The first one, a non-parametric approach, is to
divide the SFH into a number of time bins and to consider all stars
that form in each bin as an SSP. The SFH is then specified by the average
star formation rates in individual time bins.  This approach avoids the need for a
specific model for the SFH, and has been adopted in a number
of spectral fitting codes, such as STARLIGHT \citep{STARLIGHT} and pPXF
\citep{Cappellari2004}.
However, the problem with this approach is that the total number
of SSPs with non-zero weights limits the time resolution, and the
interpretation of the results are not straightforward.
The second way is to assume a functional form for SFH, which is specified by
a small number of parameters. The advantage here is that the SFH has
an infinite time resolution, and the number of free parameters is
usually small so that the model can be well constrained.
The disadvantage is, of course, that the functional form adopted may not
describe the real SFHs properly, which may lead to biased inferences.

 Since our main focus here is on early-type galaxies,
which often have relatively simple SFH, we choose the latter approach.
The SFH model considered here is represented by the
Gamma function:
\begin{equation}
\Psi(t)=\frac{1}{\tau\gamma(\alpha,t_0/\tau)} \left({t_0-t\over \tau}\right)^{\alpha-1}
    e^{-(t_0-t)/\tau}\,,
\label{gamma-sfh}
\end{equation}
where $t_0-t$ is the look-back time,
$\gamma(\alpha,t_0/\tau)\equiv \int_0^{t_0/\tau} x^{\alpha-1}e^{-x}\,dx$
is used to normalize the SFH over the age of the universe $t_0$, which is
assumed to be 14 Gyr. This is a two-parameter model,
with the parameter $\tau$ characterizing the time scale
for the onset of star formation, and the parameter $\alpha$ adopted to make
the SFH flexible. This SFH ensures that the star formation rate goes to
zero at early time, peaks at intermediate time, and then decays
exponentially at late time, as is seen for ETGs in numerical simulations
and semi-analytic models \citep{Lu2015}.

The dust attenuation is known to have degeneracy with the age and
metallicities of stars in spectral fitting. Therefore, dust extinction
has to be properly taken into account in order to derive
the properties of galaxies from their observed spectra.
In general, one assumes an attenuation curve and treats the dust
extinction as an additional model parameter. The widely used Calzetti
Law \citep{Calzetti2000} describes the dust content and opacity of
star-forming galaxies, while the two component dust model by
\cite{Charlot2000} accounts for the differences between nebular and
stellar continuum extinction. The dust properties of the ETGs
concerned here are believed to be simple, and so a single optical depth
parameter describing the attenuation of the entire stellar population
in a galaxy may be sufficient.

We note that some low-mass ETGs may contain complex stellar populations
and complex dust components, and our simple models for the SFH and dust
extinction may not work properly. For those galaxies,
a more complex model or,  alternatively, a high order
polynomials to deal with the continuum mismatch as done in pPXF,
may be needed to get better fits to their spectra. However, these
additional components often reduce the fitting efficiency and
have no straightforward interpretations. Throughout this paper,
we will stick to the simple model, but will examine the reliability
of our results against our model assumptions.

\subsubsection{Stellar kinematics}
\label{kinematics}

The intrinsic motion of stars in a galaxy induces Doppler shifts
in their spectra, and so can change the synthesized spectrum.
Detailed modeling of the stellar kinematics often uses a Gaussian-Hermite
decomposition of the line of sight velocity distribution. In practice,
however, it is usually sufficient to use the second moment of the velocity
distribution, i.e. the velocity dispersion $\sigma$, to model the
effect. Furthermore, as each of our stacked spectrum is the sum
of a relatively large number of pixels, a Gaussian broadening
is expected from the central limit theorem.
An estimate of the velocity dispersion in the stack can thus be
obtained by the accumulative effect of a set of Gaussian
functions characterized by the velocity dispersion of individual
spectra.  As described in \S\ref{stack}, the velocity dispersion,
$\sigma$, of a stacked spectrum is estimated from the quadratic mean of
the velocity dispersion of individual spectra in the stack, and this
estimate is denoted by $\sigma_{dap}$. The profile of $\sigma_{dap}$
exhibits a well-known negative gradient for early-type galaxies
\citep[e.g.][]{Emsellem2004}. We note that this estimate of the central
(the $r<0.1R_e$ bin) velocity dispersion is systematically
higher than that obtained from NSA due to the difference in
the aperture size, but the difference is typically smaller than 10\%.
In our modeling, we use $\sigma_{dap}$ to characterize the
motion of stars. When discussing the global properties, however,
we use the NSA dispersion, quoted as $\sigma_{*}$,
for the convenience in comparing with the other results.

However, the velocity dispersion estimated this way does not account
for the exact broadening of the stacked spectra. As the DAP measurements
of stellar velocities are used to correct the spectra to the rest frame,
the measurement errors can cause shifts in wavelength, leading to
an additional broadening in the stacked spectra. This effect is
particularly significant in the outer regions of galaxies where
the SNR is relative low \citep{Cappellari2017,Taniya2018}. In order
to obtain a more accurate estimate of the broadening for each stacked
spectrum, we perform spectral fitting with the software pPXF
in the range $3400-7400$ {\AA} (equivalent to MaNGA DAP) using
a set of E-MILES SSPs. pPXF can be used to perform full spectra
analysis, but here we use it only to estimate the broadening
of a spectrum. The use of template SSPs with different IMFs only
affects the $\sigma$ measurement at $\lesssim 1\, {\rm km/s}$
level (e.g. \citealt{Ferreras2013}), and so we fix the IMF to be
Salpeter in this part of our analysis. The effective velocity
dispersion, obtained this way from a stacked spectrum and referred
to as $\sigma_{ppxf}$, is then used in our fitting of the stacked
spectrum over the whole wavelength range to derive other quantities.
Because of the artificial broadening described above, $\sigma_{ppxf}$
are systematically larger than $\sigma_{dap}$ except in the central
region.

\subsection{The Bayesian approach}
\label{BIGS}

\begin{figure*}
\includegraphics[height=75mm]{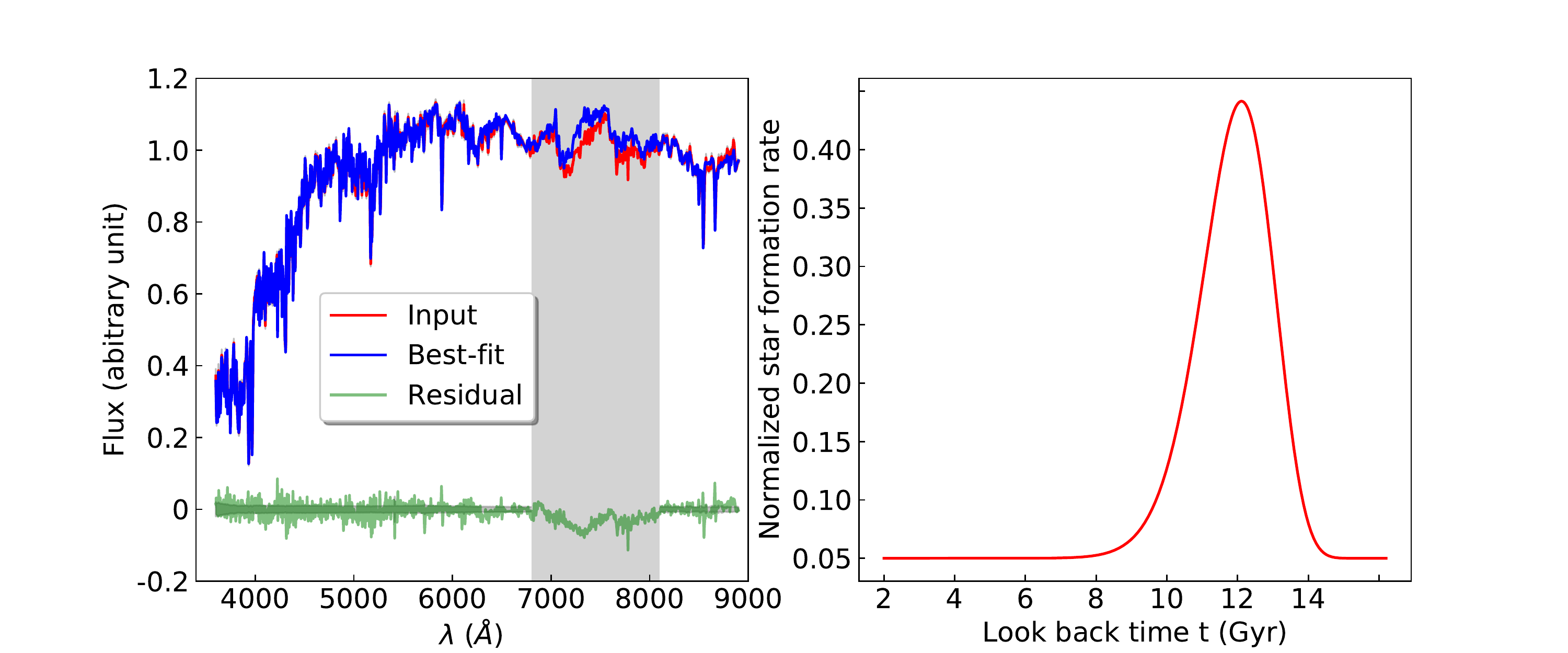}\\
\caption{Left: An example of fitting a stacked spectrum. The blue line is the
input spectrum, while the red line is the best-fit model spectrum
obtained from the posterior distribution of model parameters. The green line
shows the residual, with grey shaded regions denoting the noise level of the data. The vertical shaded
region indicates the part of the spectrum that is masked out in the fitting due to affects from residual telluric
absorption and flux calibration difficulties. Right: The corresponding best-fit star formation
history (normalized to $1 {\rm M_{\odot}}$) of the galaxy inferred from
the posterior distribution.}
\label{example-fitting}
\end{figure*}

\begin{figure*}
\includegraphics[height=180mm]{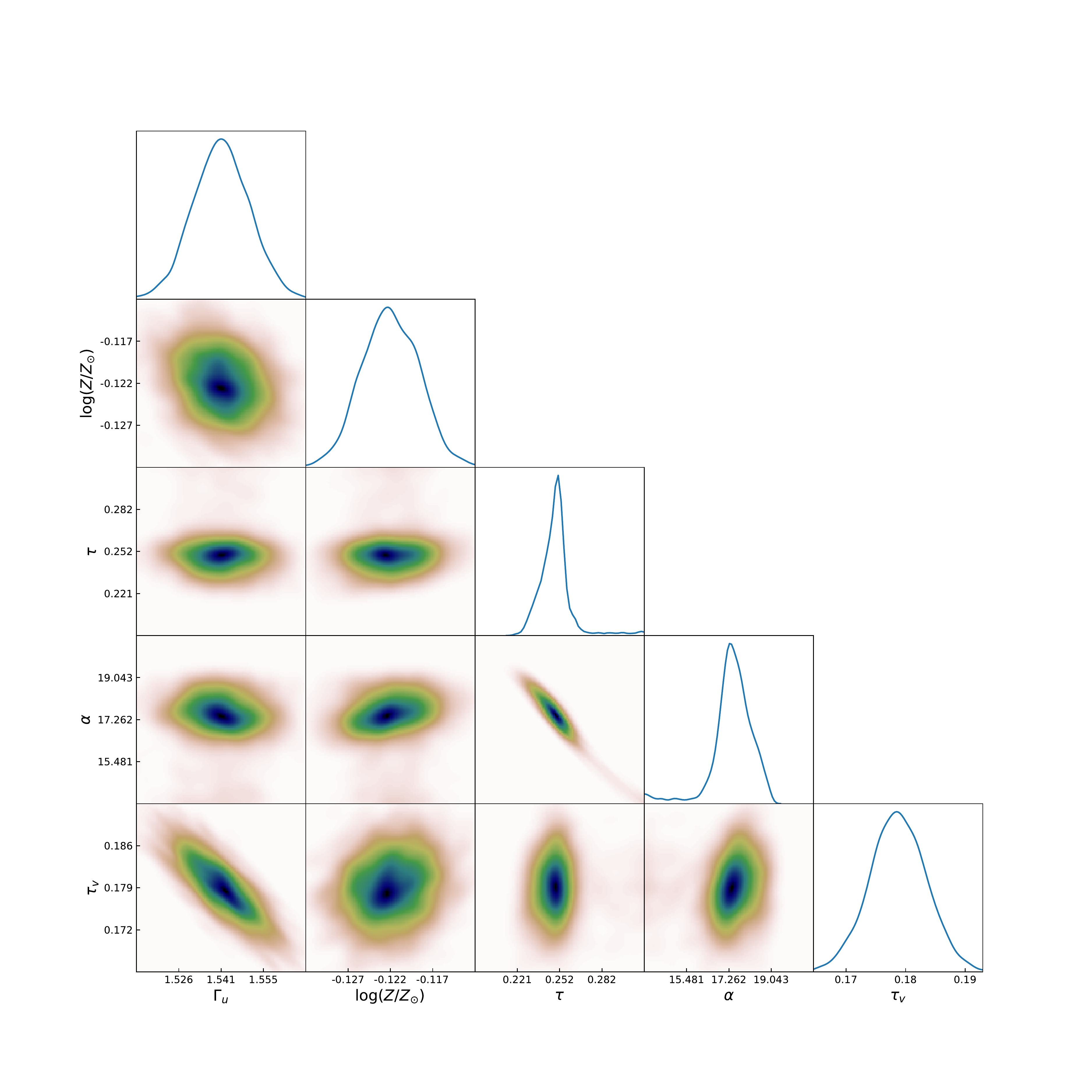}\\
\caption{Posterior probability distribution of the model parameters
listed Table \ref{tab:parameters} for the spectrum shown in Fig.\,\ref{example-fitting}.
The diagonal panels show the marginal distributions of
$\Gamma_u$, $\log(Z/Z_{odot})$, $\tau$, and $\alpha,\tau_v$,
respectively. The off-diagonal panels show the joint distributions of
parameter pairs.}
\label{example-post}
\end{figure*}
The spectral fitting code used here is called the Bayesian
Inference of Galaxy Spectra (BIGS), which can perform full Bayesian
analysis to infer physical properties of galaxies from their spectra.
The idea of Bayesian approach is to provide a statistically rigorous
way to estimate the parameters, $\theta$, that specify a model (or hypothesis),
$H$, given the constraining data $D$. Bayes' theorem states that the
probability distribution of $\theta$ given $D$ can be written as
\begin{equation}
P(\theta|D,H)=\frac{P(D|H,\theta)P(\theta|H)}{E}\,.
\end{equation}
Here $P(\theta|D,H)$ is the {\it posterior} probability distribution
of the model parameters. $P(D|H,\theta)$ is the
{\it likelihood} function, which describes the probability distribution of
the data for the model $H$ with parameters $\theta$.
$P(\theta|H)\equiv \pi(\theta)$ represents our {\it prior} knowledge about the
model parameters. The normalization,
\begin{equation}
E\equiv P(D|H) = \int P(D|H,\theta)P(\theta|H) d\theta \,,
\end{equation}
is the {\it Bayesian evidence} which describes how well the hypothesis $H$
can accommodate the data, which is of special importance in Bayesian model
selection. In the Bayesian context, the evidence ratio measures
the posterior probability ratio between two competing model families.
Suppose we have two models, $H_1$ and $H_2$, then
\begin{equation}
\frac{Pr(H_1|D)}{Pr(H_1|D)}=
\frac{Pr(D|H_1)Pr(H_1)}{Pr(D|H_2)Pr(H_2)}
=\frac{E_1}{E_2}\frac{Pr(H_1)}{Pr(H_2)}\,,
\end{equation}
where $Pr(H_1)/Pr(H_2)$ is the prior probability ratio for the two
models. If there is no a priori reason for preferring one model
over the other, we can set $Pr(H_1)/Pr(H_2)=1$, so that
the ratio of the posterior probabilities of the two models
given the data $D$ is equal to their evidence ratio.

For the problem concerned here, BIGS works as follows. To begin with,
a data spectrum, and a set of model spectra convolved with a Gaussian to
account for instrumental resolution and velocity dispersion of stars,
are provided to BIGS. BIGS then uses the Bayesian sampler, MULTINEST
\citep{Feroz2009,Feroz2013}, to produce a set of proposal parameter
vectors for the spectral synthesis model, including parameters
describing SFH, IMF, metallicity and dust attenuation. A model spectrum is generated
from these parameters, and is used to calculate the likelihood by comparing
the model prediction with the data (see eq.\ref{likelyhood} below for an example). This likelihood is then
returned to the MULTINEST sampler, which makes a decision to
accept or reject the proposal on the basis of the posterior probability
and generates a new proposal for the model. This MULTINEST-model loop
continues until convergence is achieved. The converged states, which
sample the full probability distribution of the model parameters
(the posterior distribution), are stored and to be used to derive
statistical inferences of the model.

\subsection{The fitting procedure}
\label{ssec_fittingproc}

To make quantitative statements about the IMF of a galaxy, we fit
E-MILES templates to the stacked spectra, using the procedure described
below. Our extensive test shows that almost all SSP templates have
difficulties to fit the observed continua in the wavelength range
6800-8100 {\AA}.
This mismatch between model and data results in a flux difference
 at several percent level and can potentially bias our inferences about IMF.
It is commonly found in several studies and can relate to issues
in flux calibration difficulties in the templates (e.g. \citealt{Conroya2012}),
residual telluric absorption (e.g. \citealt{Vaughan2018}), or even the
flux calibration of the data itself. The introduction of a high order
polynomial, as in MaNGA DAP, or splitting the wavelength range
to be fitted into several segments (e.g. \citealt{Vaughan2018})
can relieve the problem, but significant residuals still present.
We choose to mask out
the entire problematic region, between 6800 and 8100 {\AA},
and only to use the rest of the spectrum in our fitting.
Note that spectral regions that contain most of the IMF-sensitive
features, such as TiO1 at 5970 {\AA}, TiO2 at 6240 {\AA}, NaI doublet at 8190 {\AA},
and the calcium triplet CaT at 8400-8600 {\AA}, are used in our fitting,
although constraints on the IMF may also come from other
spectral properties.

As described in \S\ref{kinematics}, to take into account effects of stellar
kinematics, we first fit a stacked spectrum using pPXF to extract an effective velocity
dispersion, $\sigma_{\rm ppxf}$, which contains all factors that contribute to spectral
broadening, such as errors in the measurements of stellar velocities,
intrinsic dispersion, and instrumental resolution. Template spectra from the E-MILES
library are convolved with a Gaussian kernel according to this effective velocity
dispersion. We have made tests by using the NSA dispersion instead of
$\sigma_{\rm ppxf}$, and found that the measured stellar population
parameters remain stable, presumably because our measurements are not sensitive
to the details of spectral broadening. Note that, in this step of spectrum fitting,
spectra with apparent emission lines are identified, and are masked out in
subsequent analyses. However, since emission lines are not common
and generally quite weak for galaxies in our sample, excluding these lines
does not make a significant difference.

After the pre-processing described above, both the model and data spectra
are first normalized in the wavelength window $4500-5500$ {\AA}, so as to match
the stacking procedure described in \S\ref{stack}, and then sent to BIGS.
We have made tests by using normalization in different spectral regions, and
found that our results are not affected.
BIGS runs the fitting loop as described in \S\ref{BIGS}, with a flat prior
and a $\chi^2$-like likelihood function, $L(\theta)\equiv
P(D\vert H, \theta)$, which can be written as
\begin{equation}
\label{likelyhood}
\ln {L(\theta)}\propto-\sum_{i,j=1}^N\left(f_{\theta,i}-f_{D,i}\right)\left({\cal
M}^{-1}\right)_{ij}\left(f_{\theta,j}-f_{D,j}\right)\,
\end{equation}
where $N$ is the total number of wavelength bins, $f_{\theta}$ and $f_{D}$ are
the flux predicted from the parameter set $\theta$ and that of the data spectrum,
respectively, and ${\cal M}_{ij}\equiv
\langle \delta f_{D, i}\delta f_{D, j} \rangle$
is the covariance matrix of the data.
Convergence is assumed to be achieved when changes in the Bayesian evidence $\Delta \ln{E}<0.5$ in a new loop.
After convergence, the sample of the posterior of the model parameters
and the overall evidence of the model are stored for subsequent
analyses. We list all the fitting parameters in Table \ref{tab:parameters},
together with their prior distributions (assumed to be flat).
An example of the fitting is shown in Fig.\,\ref{example-fitting}
together with the inferred SFH. The posterior distributions
of the model parameters are shown in  Fig.\,\ref{example-post}.

\begin{table}
	\centering
	\caption{Priors of model parameters used to fit galaxy spectra}
	\label{tab:parameters}
	\begin{tabular}{lccr}
		\hline
		Parameter & description & Prior range\\
		\hline
		$\Gamma_u$;~$\Gamma_b$ & IMF slopes & $[0.3, 3.5]$\\
		$\log(Z/Z_{\odot})$ & metallicity & $[-2.3, 0.2]$\\
		$\tau$ & SFH parameter in Eq.\,(\ref{gamma-sfh}) & $[0.0,10.0]$\\
        $\alpha$ & SFH parameter in Eq.\,(\ref{gamma-sfh}) &$[0.0,20.0]$\\
        $\tau_v$ & dust optical depth at 5500 \AA & $[0.0,2.0]$\\
		\hline
	\end{tabular}
\end{table}

%
%
\section{Inferences of the variations of the initial mass function}
\label{result}

\subsection{IMF model selection}

\begin{figure}
\includegraphics[height=70mm]{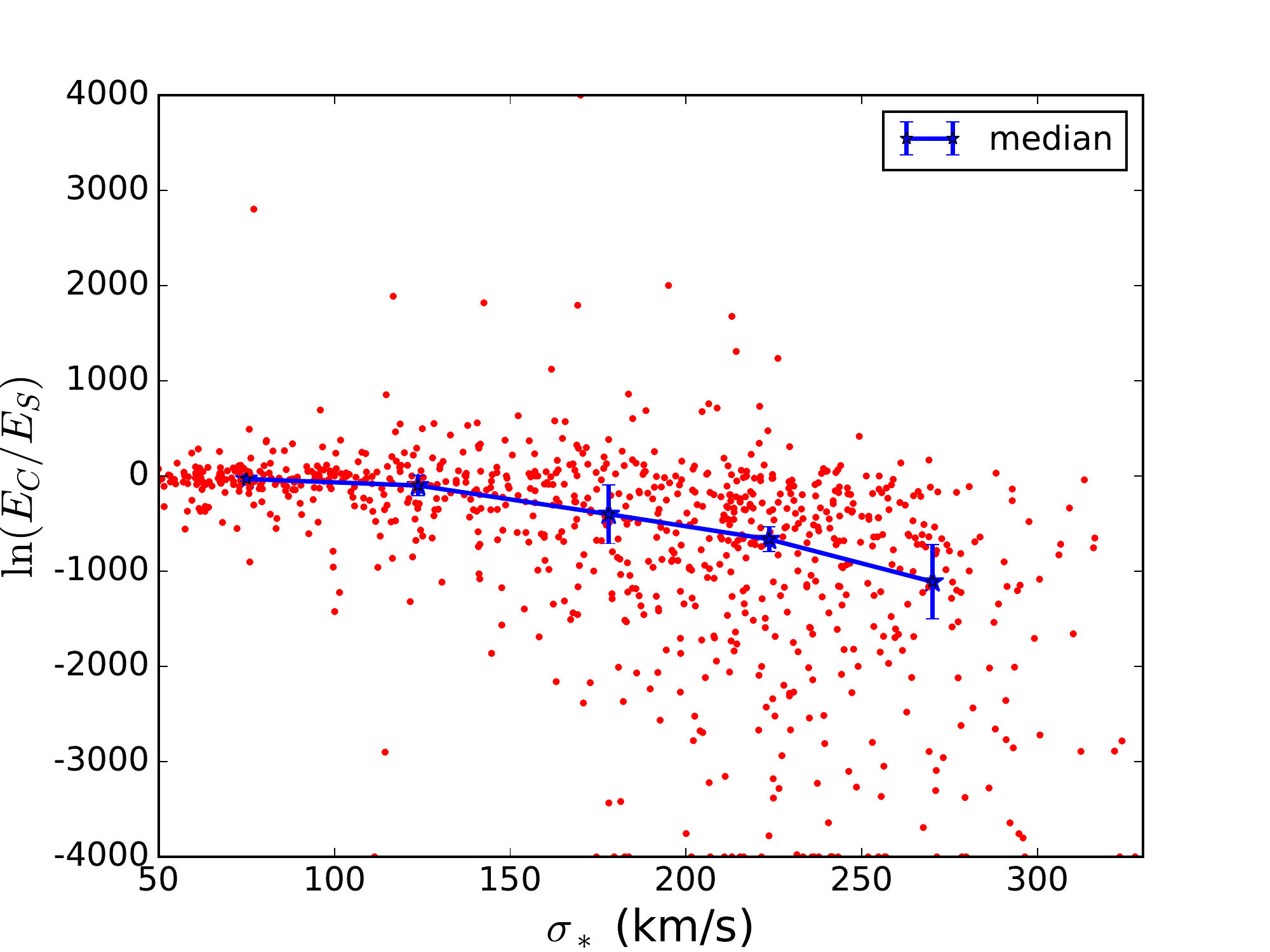}\\\
\caption{The evidence ratio between Chabrier and Salpeter IMF as a function of
galaxy velocity dispersion. Each red dot stands for the result of a MaNGA
ETG. Blue stars are the median values in five $\sigma_*$ bins and are
linked by a blue line. Error bars are  obtained from the jackknife resampling method.
}
\label{evratio}
\end{figure}

\begin{figure}
\includegraphics[height=70mm]{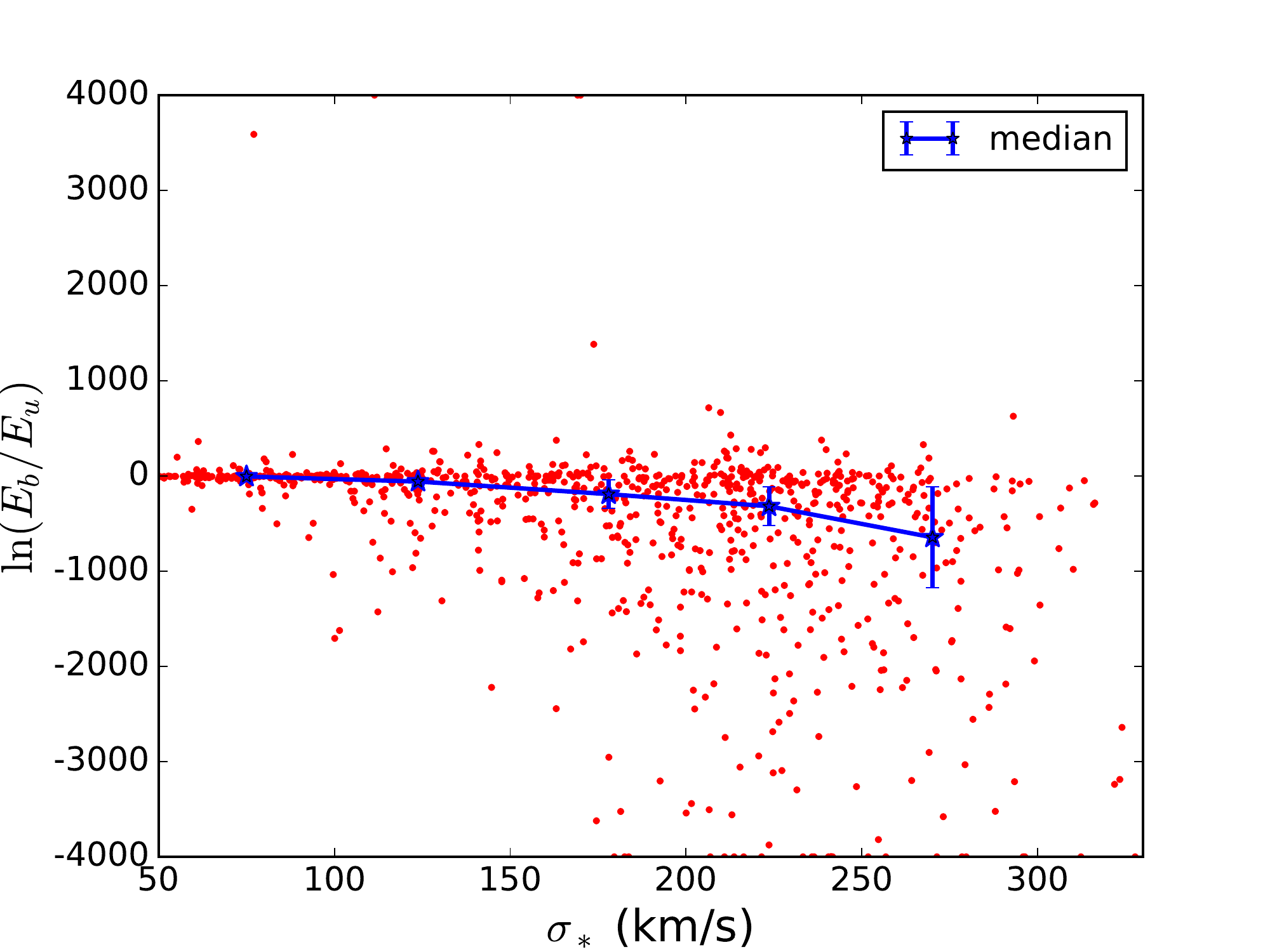}\\\
\caption{The evidence ratio between bimodel and unimodel as a function of
galaxy velocity dispersion. Each red dot stands for the result of a MaNGA ETG.
Blue stars are the medians in the five $\sigma_*$ bins and linked by a
blue line. Error bars are obtained from the jackknife resampling method.
}\label{evratio_bi}
\end{figure}

The Bayesian approach described above can provide an overall
evidence for any given model family of the IMF. As described in \S\ref{BIGS},
the Bayesian evidence ratio is a measure of the posterior probability
ratio between two competing model families and, therefore, can be used
to distinguish different models. Thus, before investigating the
variations of the IMF with galaxy properties, it is useful to use Bayesian
evidence to examine whether the observational data prefers one form of the
IMF among all the forms described above. To this end, we
first fit the stacked spectra (each being a stack of pixels
within the effective radius of a galaxy)
of individual galaxies with the two
selected set of E-MILES SSPs assuming the two widely-used IMFs, the Chabrier IMF
and the Salpeter IMF, while keeping all other parts of the model
intact. We then use the evidence ratio between the two fits to gauge
the preference of the data between the two models. Fig. \ref{evratio}
shows $\ln(E_{\rm C}/E_{\rm S})$ versus the NSA velocity dispersion,
$\sigma_*$, obtained from spatially unresolved analysis,
where $E_{\rm C}$ and $E_{\rm S}$ stand for the Bayesian evidences of
the Chabrier and Salpeter IMF models, respectively. The median values
in five $\sigma_*$ bins are plotted as blue stars to demonstrate the average trends.
To estimate the statistical error, we randomly divide
our sample into 20 sub-samples, and calculate the median in 20 different jackknife
copies by eliminating one of the 20 sub-samples. The statistical error is
then derived from the 20 jackknife medians, which are shown in
Fig. \ref{evratio} as error bars. There is a clear trend in the evidence
ratio with the velocity dispersion: the evidence ratio decreases with increasing
$\sigma_*$. Overall, there is a preference to the Salpeter
IMF by galaxies with $\sigma_*>150\,{\rm km/s}$, while
such a preference is absent for galaxies with lower
$\sigma_*$. Note that very few galaxies of high $\sigma_*$
prefer Chabrier IMF over Salpeter IMF, although
many of them appear indifferent to both.

Next we compare the two model families, unimodel and bimodel.
Fig. \ref{evratio_bi} shows $\ln(E_{\rm b}/E_{\rm u})$ versus $\sigma_*$,
where $E_{\rm b}$ and $E_{\rm u}$ stand for the Bayesian evidences of
the bimodel and unimodel IMFs, respectively. Interestingly, although
results are generally concentrated around zero, there is a
weak trend showing a decreasing evidence ratio with increasing
$\sigma_*$, i.e. a preference to the unimodel IMF. This result
indicates that the IMF of early type galaxies is more
likely to be a power law all the way to the low mass end,
instead of becoming flat as does the Galactic IMF, such as
the Chabrier IMF \citep{Chabrier2003} and the Kroupa IMF \citep{Kroupa2001}.
This result is, however, in conflict with some of the results that take into account dynamical constraints \citep[e.g.][]{Barbera2013,Lyubenova2016}, which indicates that the IMF parameterization
adopted here may not be a good model for these galaxies
(see discussions in \S \ref{sec_comparisons_slope}).

Unfortunately, it is not straightforward to have a quantitative
decision about the IMF type. A naive classification, such as
identifying all galaxies with $\ln(E_C/E_S)>0$ as the ones
that prefer the Chabrier IMF, may not be unreliable,
because the scatter presented in Fig.\,\ref{evratio}
is large and it is unclear whether it is produced by
intrinsic variations or by noise. In what follows,
we will present results mainly based on unimodel,
although we sometimes also show bimodel results
for comparison.

\subsection{Dependence on galaxy velocity dispersion}
\label{gammavsvd}

\begin{figure*}
\includegraphics[height=70mm]{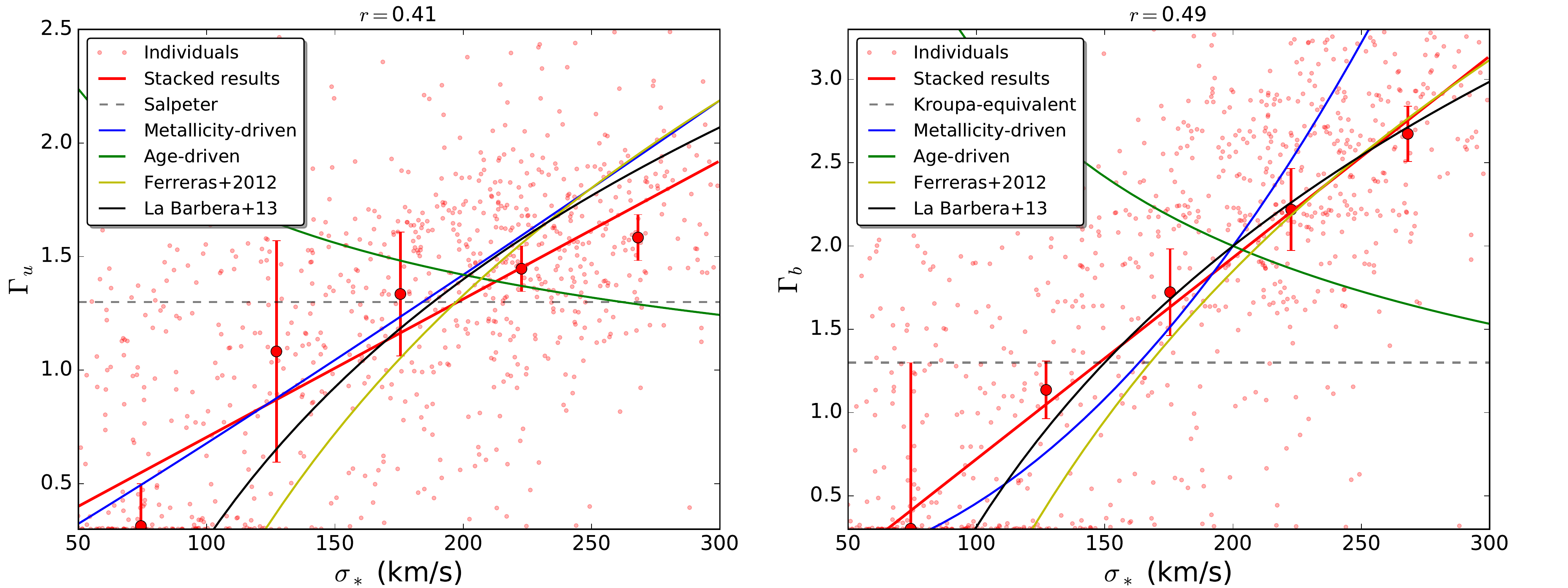}\\
\caption{The best-fit slope of the unimodel (left) and bimodel (right) IMF for MaNGA ETGs plotted
against central velocity dispersion. In each panel, small red dots
stand for the results of individual MaNGA ETGs.
The Pearson correlation coefficient, $r$, is
shown on the top of each plot to quantify the correlation between the
IMF slope and the velocity dispersion.
 Big thick dots are the results
derived from the stacked spectra in five $\sigma_*$ bins,
while the red line is a fit to these points. The errorbars are obtained from
20 jackknife samples. The grey dash line denotes the slope of the Salpeter IMF.
Blue and green lines are predictions by assuming the IMF variation is
is driven by metallicity and age (see \S \ref{ssec_drive}), respectively.
Results from \citet{Ferreras2013} and \citet{Barbera2013} are shown as
yellow and black lines, respectively.
}\label{un}
\end{figure*}

\begin{figure}
\includegraphics[height=70mm]{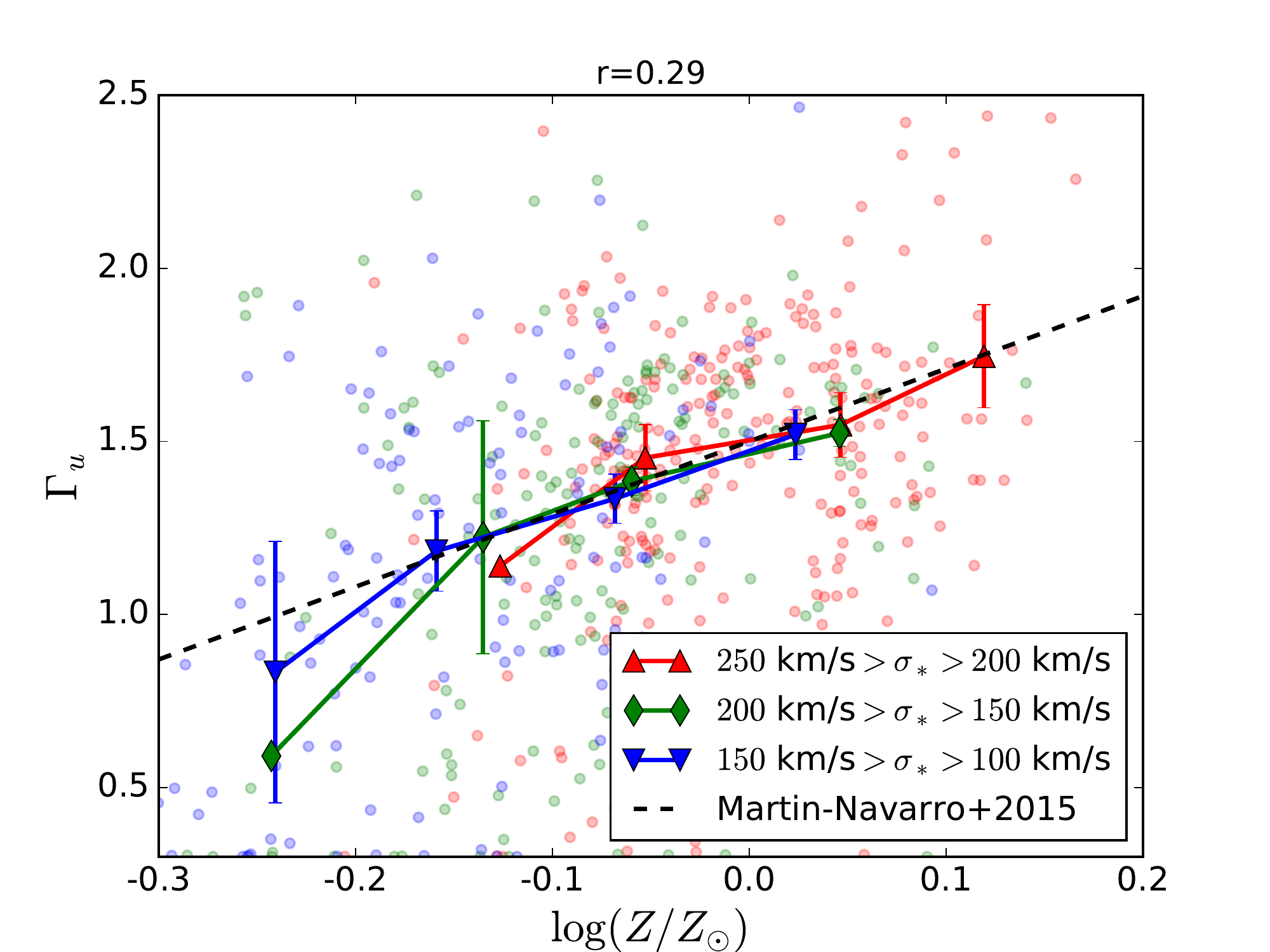}\\
\caption{Unimodel IMF slope as a function of metallicity in different $\sigma_*$ bins.
The red, green, blue dots stand for galaxies in three $\sigma_*$ bins, as indicated.
The Pearson correlation coefficient, $r$, between the IMF slope and metallicity
is shown on the top of the panel.
The big symbols connected thick lines are the median values in each
$\sigma_*$ bin. Error-bars are obtained from the jackknife method.
The black dash line shows the result from \citet{Navarroa2015}.
}
\label{sigmainmetal}
\end{figure}

\begin{figure}
\includegraphics[height=70mm]{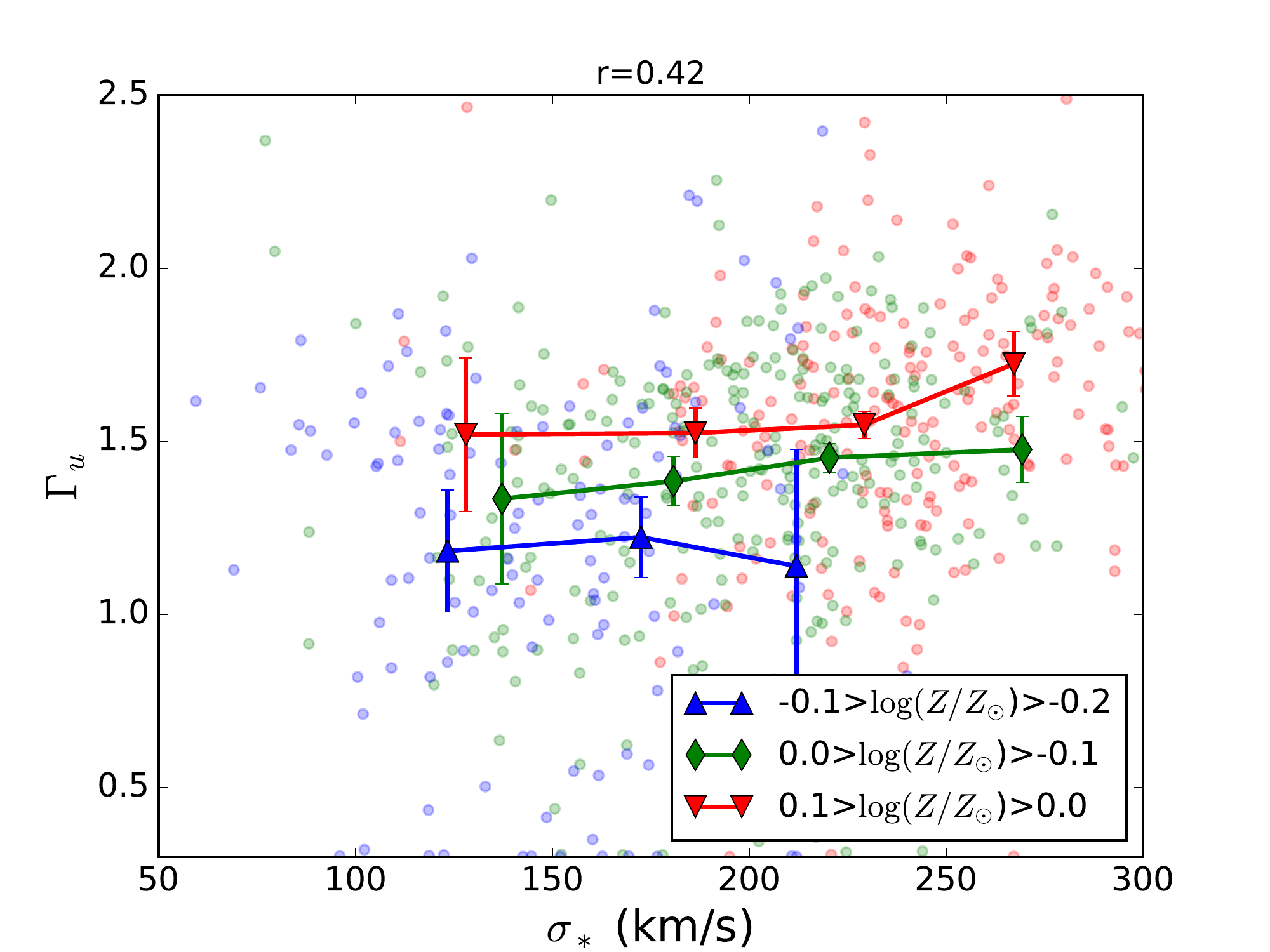}\\
\caption{Unimodel IMF slope as a function of $\sigma_*$ in different metallicity bins.
The red, green, and blue dots stand for galaxies in three
metallicity bins, as indicated.
The Pearson correlation coefficient, $r$,
between $\Gamma_u$ and $\sigma_*$ is shown on the
top of the panel.
The big symbols connected by thick lines are median values in each $Z$ bin.
Error-bars are obtained from the jackknife method.
}
\label{metalinsigma}
\end{figure}

\begin{figure*}
\includegraphics[height=37.5mm]{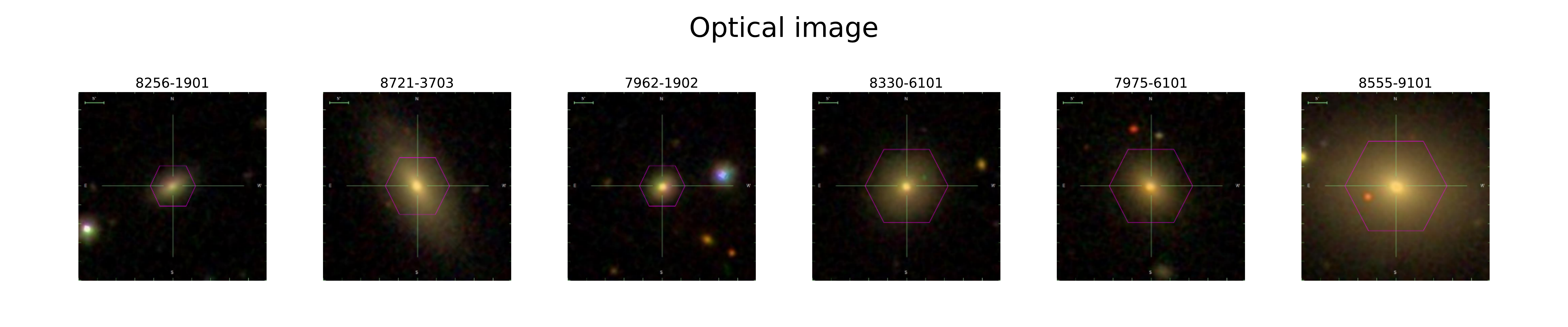}\\
\includegraphics[height=90mm]{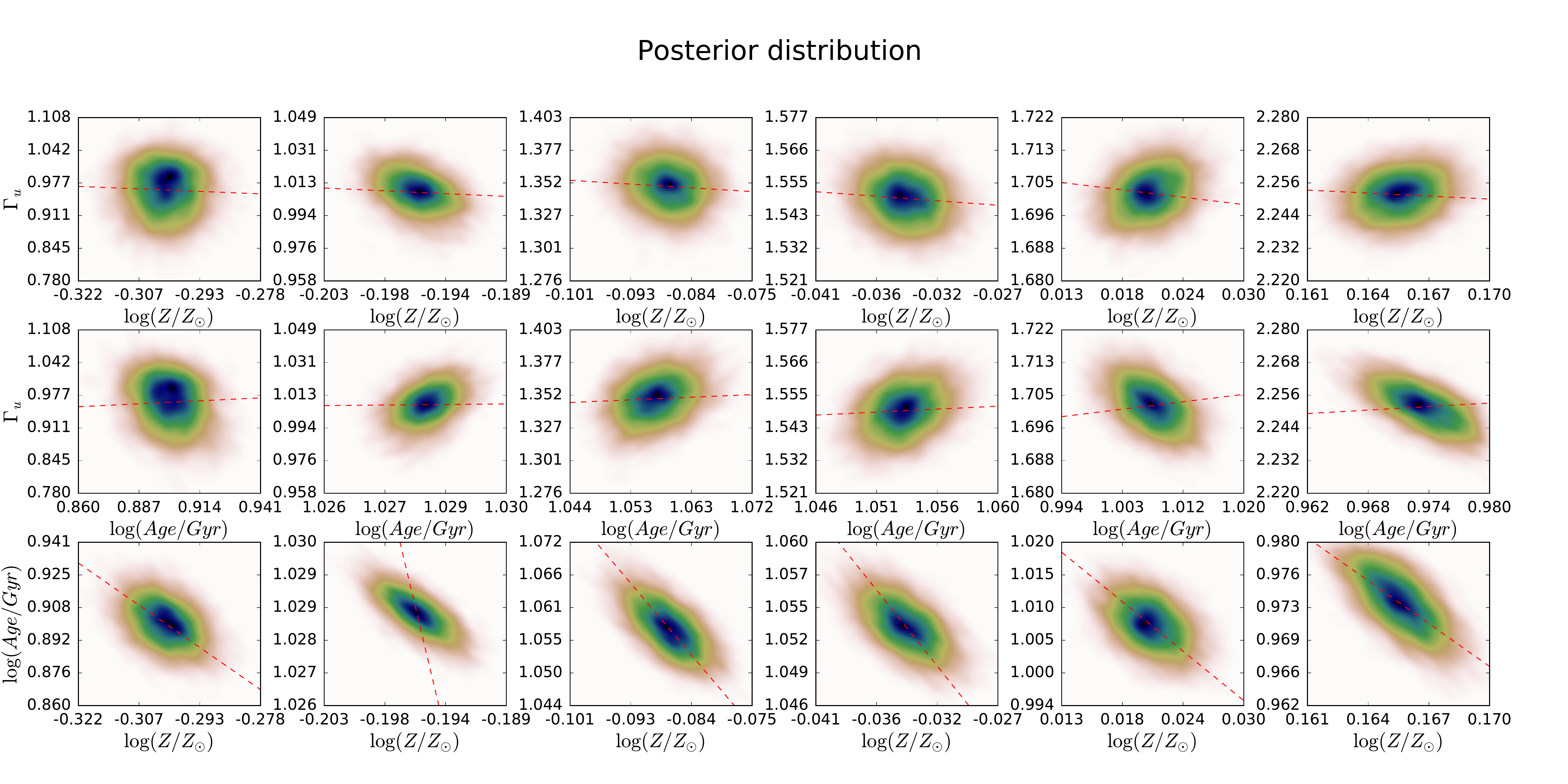}\\
\caption{Top panels: optical images of six example galaxies, with
MaNGA plate-ifu shown on top of each image.
Bottom panels: joined posterior distribution estimated from the fitting results
with standard setting. The first and second rows show the
covariance between the IMF slope, $\Gamma_u$, with the stellar
metallicity, $Z$, and mean stellar age, $A$, respectively.
The last row shows the covariance between $A$ and  metallicity $Z$.
Red dash lines in each row show the degeneracy line inferred from
the median value of the whole sample. The slopes of these lines are
$\Delta\Gamma_u/\Delta\log(Z)=-0.34$, $\Delta\Gamma_u/\Delta\log(A)=0.22$,
and $\Delta\log(A)/\Delta\log{Z}=-1.42$, respectively.
}
\label{degen-sample}
\end{figure*}

We fit the MaNGA spectra (stacked within $1 R_e$) of individual galaxies
with both unimodel and bimodel described above.
The Gamma model for the SFH (referred to as $\Gamma$-SFH)
and a one-component dust model are assumed, together with a
metallicity parameter. The fits to the spectra are in general quite good,
with residual at around the $3\%$ level over a large range of
wavelength. The results for unimodel and bimodel
are shown in  Figs.\,\ref{un}. The best-fit slopes of the IMF are generated
from the posterior distribution through marginalizing over the rest of model
parameters. The velocity dispersion used in the plots is the NSA
velocity dispersion, $\sigma_*$, adopted for the convenience
to compare with previous studies.
The small red dots are the results for individual galaxies, while
the bigger blue squares are the mean results in five $\sigma_*$ bins,
obtained by fitting the total composite spectrum in each of the
$\sigma_*$ bins. We again make use of the jackknife resampling method
to estimate the statistical error. To this end, we randomly divide
our sample into 20 sub-samples, and generate 20 different jackknife
copies of the stacked spectra by eliminating one of the 20 sub-samples
in each stack. The same fitting algorithm is then applied to each
of the 20 jackknife copies to obtain the corresponding IMF slope. The
variance of the IMF slope values estimated from the jackknife copies
are shown in the in Figs.\,\ref{un} as the error bars.
The Pearson correlation coefficient $r$ between the pair
of plotted quantities is shown on the top of each panels to
quantify the strength of the correlation.

As one can see, the IMF slopes in both unimodel and bimodel
increase with the velocity dispersion of the ETGs.
This indicates clearly that the IMF is non-universal, but changes
systematically in that the contribution of low-mass stars is
higher in more massive galaxies. In the unimodel, the IMF
slope can even be steeper than the Salpeter IMF for ETGs
with the highest $\sigma_*$, suggesting that the IMF of these
galaxies is significantly bottom-heavy. If the IMF is forced
to have a turnover at $m<0.6$, as in bimodel, the majority
of the galaxies at $\sigma_*>200\,{\rm km/s}$ require
$\Gamma_b>1.35$, again indicating the preference to a
bottom-heavy IMF. Note that $\Gamma_b$ has stronger
dependence on $\sigma_*$ than $\Gamma_u$, because the
suppression of stars at $m<0.6$ in bimodel has to be
compensated by a steeper slope so as to produce sufficient
amounts of low-mass stars to match the spectra of high
$\sigma_*$ galaxies. As shown in the plot, our results
are in good agreement with those of \cite{Ferreras2013} and
\cite{Barbera2013}, which is expected as we are using similar
stellar population models although different fitting approaches.

\subsection{Dependence on metallicity}

The small points in Fig.\,\ref{sigmainmetal} show the IMF slope, $\Gamma_u$,
against galaxy metallicity for individual galaxies. Here the metallicities are
obtained directly from our fitting to individual spectra and marginalizing
over the posterior distributions. We focus on galaxies with
metallicities in the range $-0.3<\log(Z/Z_{\odot})<0.2$ in which the
majority of our sample galaxies are contained.
As one can see, there is a strong correlation between
the IMF shape with metallicity. This correlation
may not be surprising, as galaxy metalicities are known to
be correlated with the depth of its gravitational well,
as represented by its velocity dispersion \citep[e.g.][]{Gallazzi2005}.
Our result is in good agreement with that of \cite{Navarroa2015}, who
found that the correlation of the IMF shape with metallicity
is the strongest among those with a number of other
galaxy properties, including galaxy velocity dispersion.
However, as can be seen from the correlation
coefficients, the IMF slope-metallicity relation
is not as tight as the IMF slope-$\sigma_*$ relation.
This is likely due to age-metallicity degeneracy,
which has not been taken into account here but will be
discussed in more detail in \S \ref{ssec_drive}.

The large sample used here allows us to analyze the
correlations of the IMF shape with velocity dispersion and
metallicity independently, so that we can investigate whether
the systematic change of the IMF slope is driven by
velocity dispersion or by metallicity. The solid lines in
Fig.\,\ref{sigmainmetal} show the average
$\Gamma_u$ - metallicity relations in three $\sigma_*$
bins. As one can see, the dependence on metallicty
is very similar for galaxies of different velocity
dispersions, indicating that it is the matallicity
that produces the $\sigma_*$ dependence of the IMF slope.
As comparison, we show in Fig.\,\ref{metalinsigma} the IMF
slope as a function of galaxy velocity dispersion for
individual galaxies separated into different
metallicity bins. The dependence of the IMF slope on $\sigma_*$
is significantly weakened for galaxies of similar metallicities,
suggesting again that the dependence on $\sigma_*$ is metallicity driven.

Since both the IMF slope and metallicity are model parameters
in our spectrum fitting, it is important to check if the
dependence of $\Gamma_u$ on $Z$ is produced by the degeneracy
of these two parameters in the model. Our Bayesian approach
allows us to examine such degeneracy for individual galaxies.
In Fig.\,\ref{degen-sample} we show the posterior distributions
obtained from the MULTINEST sampling for six representative galaxies
with different stellar masses, metallicities, and spectral
signal-to-noise ratios. As one can see from the first row in
the lower part of the figure, the statistical inference errors
in both $\Gamma_u$ and $Z$ are much smaller than the ranges over which
the correlation between $\Gamma_u$ and $Z$ is observed,
indicating that the model parameters for individual
galaxies are stringently constrained by their spectra.
In particular, the median degeneracy relation obtained from
the total sample, which has a slope
$\Delta\Gamma_u/\Delta\log(Z)=-0.34$, as represented by the red
dashed lines, is quite weak, indicating that the
dependence of the IMF slope on $Z$ is not produced by the degeneracy
of the two parameters in our spectrum fitting.

\subsection{Dependence on stellar age and dust extinction}

\begin{figure}
\includegraphics[height=70mm]{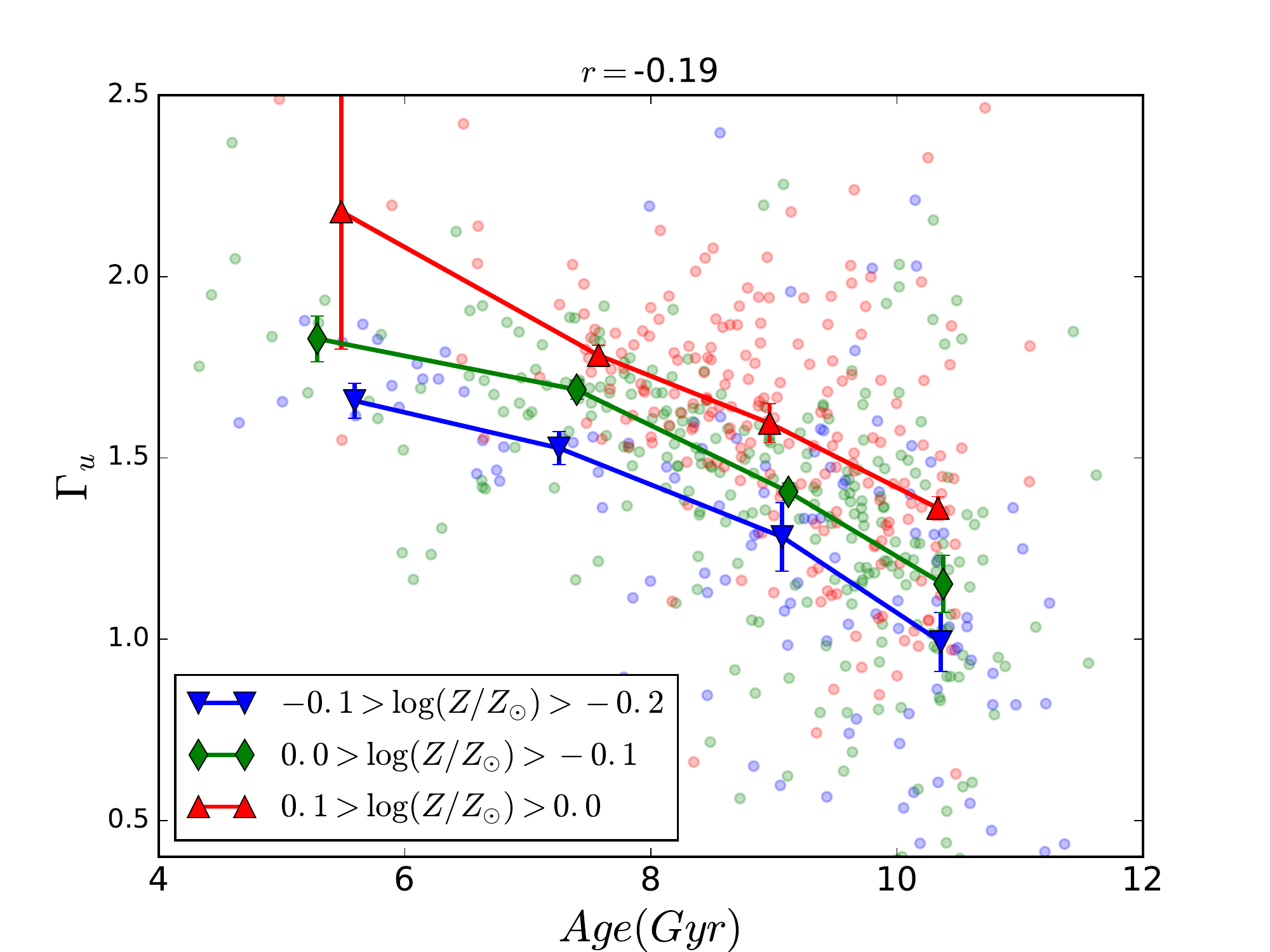}\\
\caption{Unimodel IMF slope as a function of light weighted age.
The red, green, and blue dots stand for galaxies in three
metallicity bins, as indicated. The Pearson correlation coefficient
$r$ between $\Gamma_u$ and age is shown on the top of the plot.
The big symbols
connected by thick lines are median values in each metallicity bin.
Error bars are based 20 jackknife samples.}
\label{metalinage}
\end{figure}

\begin{figure*}
\includegraphics[height=70mm]{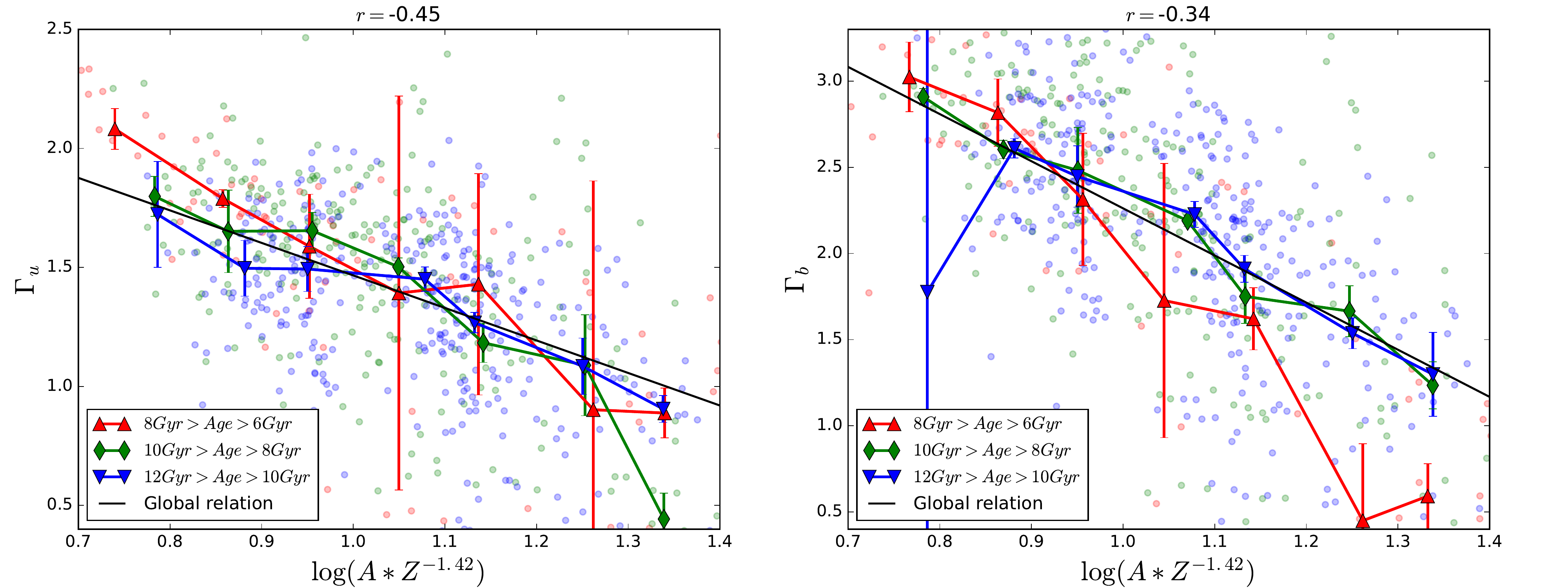}\\
\caption{Unimodel (left) and bimodel (right) IMF slope as a function of $D\equiv AZ^{-1.42}$, where
the age, $A$,  is in units of Gyr and metallicity, $Z$,
in units of solar metallicity. In each panel, the red, green, and blue dots
are for galaxies in three age bins, as indicated.The Pearson correlation
coefficient, $r$, between the quantities plotted, is shown on the top of each plot.
The big symbols connected by thick lines are median values in each age
bin. Error bars are based on 20 jackknife samples. The black straight lines
show the fits to all the data points, and are
$\Gamma_u \propto D^{-1.37}$ and $\Gamma_b \propto D^{-2.73}$,
respectively.
} \label{unvsD}
\end{figure*}

 From the constrained star formation history of a galaxy,
we can also estimate a light-weighted average of stellar ages,
$A$, to represent the characteristic age of the galaxy.
The light-weighted ages are estimated as follows.
With each given pair of $\tau$ and $\alpha$ values,
Eq.\,(\ref{gamma-sfh}) gives the normalized star formation rate
as a function of time. Thus, the total mass of stars that
form in a time interval at $t_i$ can be written as
$\Delta M_i={\rm SFR}(t_i) \Delta t_i$.
Since the fluxes of SSPs provided by stellar population
models in BIGS are normalized to $1 M_{\odot}$ \citep{Vazdekis2016},
we assign $\Delta M_i$ to the SSP of the corresponding age and calculate
a final spectrum as the sum of the fluxes from these SFR-weighted SSPs.
In contrast, in traditional algorithms based on discrete
SSPs, SSP fluxes from stellar population models are normalized at a specific
wavelength $\lambda_0$. Light fractions are then assigned to such
normalized SSP and the final spectrum is then calculated
from the sum of the fluxes from these light-weighted SSPs \citep{STARLIGHT}.
To calculate a similar light-weighted age, we calculate the light fraction of
the SSP that forms in the $i$th time interval, $x_i$, from the
stellar mass of this SSP, $\Delta M_i$, using
$x_i=\Delta M_i f_i (\lambda_0)$, where $f_i (\lambda_0)$ is the
flux of the model SSP at the wavelength $\lambda_0$.
The light-weighted ages are then obtained as
 \begin{equation}
 \langle A\rangle ={\sum{x_i A_i}\over \sum{x_i}},
 \end{equation}
 where $A_i$ is the age for the SSP that forms in the $i$th time interval.

The small points in Fig. \ref{metalinage} show the IMF slope,
$\Gamma_u$, as a function of galaxy age, $A$,
for individual galaxies, while the big symbols connected
by solid lines are the median relations for galaxies in
different metallicity bins. The IMF shape appears to
decrease with stellar age in all metallicity bins, indicating that
the IMF slope may also depend stellar age, although the global correlation is quite weak,
as indicated by the correlation coefficient. This result is similar to the anti-correlation
between IMF slope and stellar age found in \cite{Navarroa2015} from CALIFA galaxies.
The posterior distribution between the IMF slope and stellar age for
the six representative galaxies are shown in the second row
of the lower part of Fig.\,\ref{degen-sample}. The dashed red
lines, which have a slope $\Delta\Gamma_u/\Delta\log(A)=0.22$,
show the madian degeneracy line between $\Gamma_u$ and $A$
obtained from the total sample. The small, positive slope
indicates that the dependence of $\Gamma_u$ on $A$ shown in
Fig.\,\ref{metalinage} is not produced by the degeneracy.
However, since the stellar age and metallicity are degenerate in spectrum
fitting, it is important to check if the age-dependence
of the IMF slope is independent of the metallicity-dependence.
To this end, we show in the last row of Fig.\,\ref{degen-sample}
the posterior joint distribution of $A$ and $Z$ for the representative
galaxies. For reference, we plot a dash line in each panel to show the
degeneracy line, $A Z^{\beta}={\rm constant}$, where $\beta= 1.42$, as is
obtained by the median $A$-$Z$ degeneracy relation of
individual galaxies. Note that the $\beta$ value obtained here is close to
its canonical value, $1.5$, as obtained by \citet{Worthey1994}.
There are variations in the degeneracy of $A$ and $Z$ from
galaxy to galaxy, but the trend that $A$ varies inversely
with $Z$ is the same for almost all galaxies. Because of
this age-metallicity degeneracy, it is difficult to separate the
age and metallicity effects on IMF, unless the age and metallicity
can be determined independently. We will come back to this in
\S\ref{ssec_drive}.

\begin{figure}
\includegraphics[height=70mm]{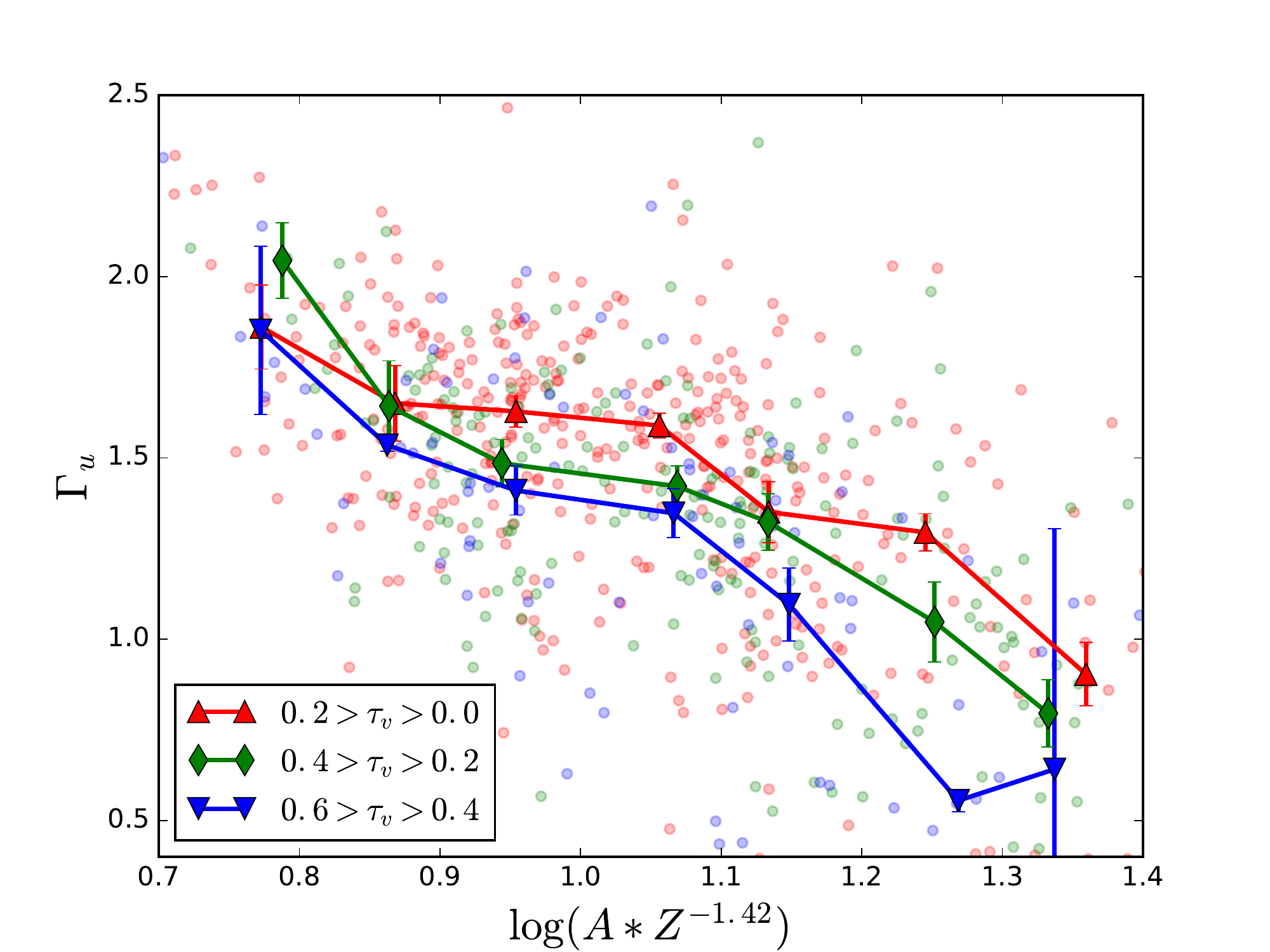}\\
\caption{
Unimodel IMF slope as a function of $D\equiv AZ^{-1.42}$, where
the age, $A$,  is in units of Gyr and metallicity, $Z$,
in units of solar metallicity.
The red, green, blue dots stand for galaxies in three bins
of the dust optical depth parameter, as indicated. The big symbols
connected by thick lines are median values in each $\tau_v$ bin.
Error bars are based 20 jackknife samples.
} \label{tauvinD}
\end{figure}

In our spectral fitting, dust extinction is modeled by a given extinction
curve, and specified by an optical depth parameter.
The dust optical depths obtained for individual galaxies
from our fitting range from 0.0 to 0.6, indicating that dust extinction
is quite small, as is expected for the ETGs concerned here.
Since the dust absorption is stronger at shorter wavelengths,
dust extinction makes a galaxy look redder, or older in its
continuum shape. Degeneracy between the dust optical depth
and the stellar age (or metallicity) is thus expected in the
spectral fitting. We examine the influence of dust in the
$\Gamma_u$-$D$ relation by separating galaxies
into three specific optical depth bins, as shown in Fig \ref{tauvinD}.
The figure shows that galaxies with higher dust optical depth
have slightly lower $\Gamma_u$. This is expected from
the degeneracy between the optical depth and age:
an overestimate of dust extinction lead to an underestimate of
stellar age (smaller $D$). The $\Gamma_u$-$D$ correlation would
then shift to the left. A possible way to take into account
this degeneracy is to use a combination of $A$, $Z$ and $\tau_v$
that is perpendicular to the degeneracy plane in the
$(A, Z, \tau_v)$ space. However, current
data is too poor to establish such a degeneracy plane.
Since the dust extinction is relatively small in our galaxies,
and its influence on the $\Gamma_u$-$D$ relation is
only modest, as shown in Fig \ref{tauvinD}, we will not
discuss the dust effect any further.

\subsection{The origin of IMF variation}
\label{ssec_drive}
In order to see the effect of the age-metallicity degeneracy
described above, we plot $\Gamma_u$ and $\Gamma_b$
versus $D\equiv AZ^{-\beta}$ in Fig.\,\ref{unvsD}.
Note that the line of a constant $D$ is perpendicular to the degeneracy
line in the $\log(A)$ - $\log(Z)$ space, so that the effects of
the age-metallicity degeneracy are expected to be
reduced or eliminated along $D$, if the age-metallicity degeneracy
were accurately described by the mean relation we use.As one can see from Fig.\, \ref{unvsD}, the scatter in
$\Gamma_u$ is indeed reduced relative to that in
the $\Gamma_u$ - $A$ and $\Gamma_u$ - $Z$ relations.
This is also true for the results of $\Gamma_b$ shown
in the right panel.
In addition, the correlation between
$\Gamma$ and $D$ is comparable to or even
slightly stronger than the $\Gamma$-$\sigma$
correlation, as seen from the correlation coefficients.
  Moreover, galaxies of
  different ages (and metallicties, not shown) now obey similar relations, indicating that the
dependencies of the IMF slope on age and metallicity are not
independent, but through the combination $D$ that is determined
by the age-metallicity degeneracy.

In order to pin down the intrinsic driver of the IMF variation,
it is necessary to determine the age and metallicity in a way such
that the degeneracy between the two is eliminated or reduced.
Unfortunately, our method alone is not able to achieve this goal.
The age-metallicity degeneracy is a well-known problem in full
spectrum analysis. The effects of such a degeneracy is seen in the
IMF analysis of \citet{Ferreras2013} based on the  E-MILES SSP
model. The anti-correlation between the IMF slope and stellar age found
for CALIFA galaxies by \citet{Navarroa2015} may suffer from the same
issue. One way to tackle the problem is use ages and metallicities
obtained from methods that can break the age-metallicity degeneracy.

By using a set of stellar population models that accounts for element
abundance ratio effects, \cite{Thomas2005} analyzed a sample of 124 early-type
galaxies and obtained scaling relations between age, metallicity and velocity dispersion.
Here we use these scaling relations as ancillary results in our analysis,
under the assumption that they are valid for our sample.
Note that the models used in \cite{Thomas2005}, originally presented
in \cite{Thomas2003}, have systematic differences from the the E-MILES models.
For example, the ages in \cite{Thomas2005} are estimated with SSP-based
algorithms in the absence of IMF variations, and the treatment of the
abundance pattern is also different. To ensure the validity of the
assumption, we have used similar age measurements and tested
that including IMF variations does not cause any systematical
changes in the age determinations. We discuss effects of abundance
variations in \S\ref{ssec_abundence}.
However, One should keep in mind that this assumption is
not extensively tested, and so the results based on it
should be treated with cautions.
Under this assumption, if the IMF variation with $D$ is totally driven
by the change of metalllicity, we can
set $D\propto Z^{-1.42}$, where $Z$ is the `intrinsic' metallicity
of the galaxy, which is not affected by the age-metallicity degeneracy.
Then the average correlations between the IMF slope and $D$
(i.e. $\Gamma_u \propto D^{-1.37}$ and $\Gamma_b \propto D^{-2.73}$, as shown
in Fig.\, \ref{unvsD}) would give $\Gamma_u \propto Z^{1.95}$ and
$\Gamma_b \propto Z^{3.88}$, respectively.
Using the scaling relations in equation (1) of \cite{Thomas2005},
$[Z/H]\propto \sigma_*^{0.55}$, we get $\Gamma_u  \propto \sigma^{1.07}$ and
$\Gamma_b  \propto \sigma_*^{2.13}$, which are shown in Fig.\,\ref{un} (blue lines)
for comparisons. One can see that these scaling relations are
consistent with the observational data. Similar scaling relations can also be made
by assuming that the IMF variation is totally driven by the change of age.
Using $\log(A)\propto \sigma_*^{0.238}$ as given in \cite{Thomas2005}, we obtain
$\Gamma_u  \propto \sigma_*^{-0.33} $ and  $\Gamma_b  \propto \sigma_*^{-0.66}$
(shown by green lines in Fig.\,\ref{un}). This age-driven model is clearly in conflict
with the observational data. These scaling results suggest that the observed IMF variation
is likely driven by metallicity, instead of age.

\subsection{Radial dependence}

\begin{figure}
\includegraphics[height=70mm]{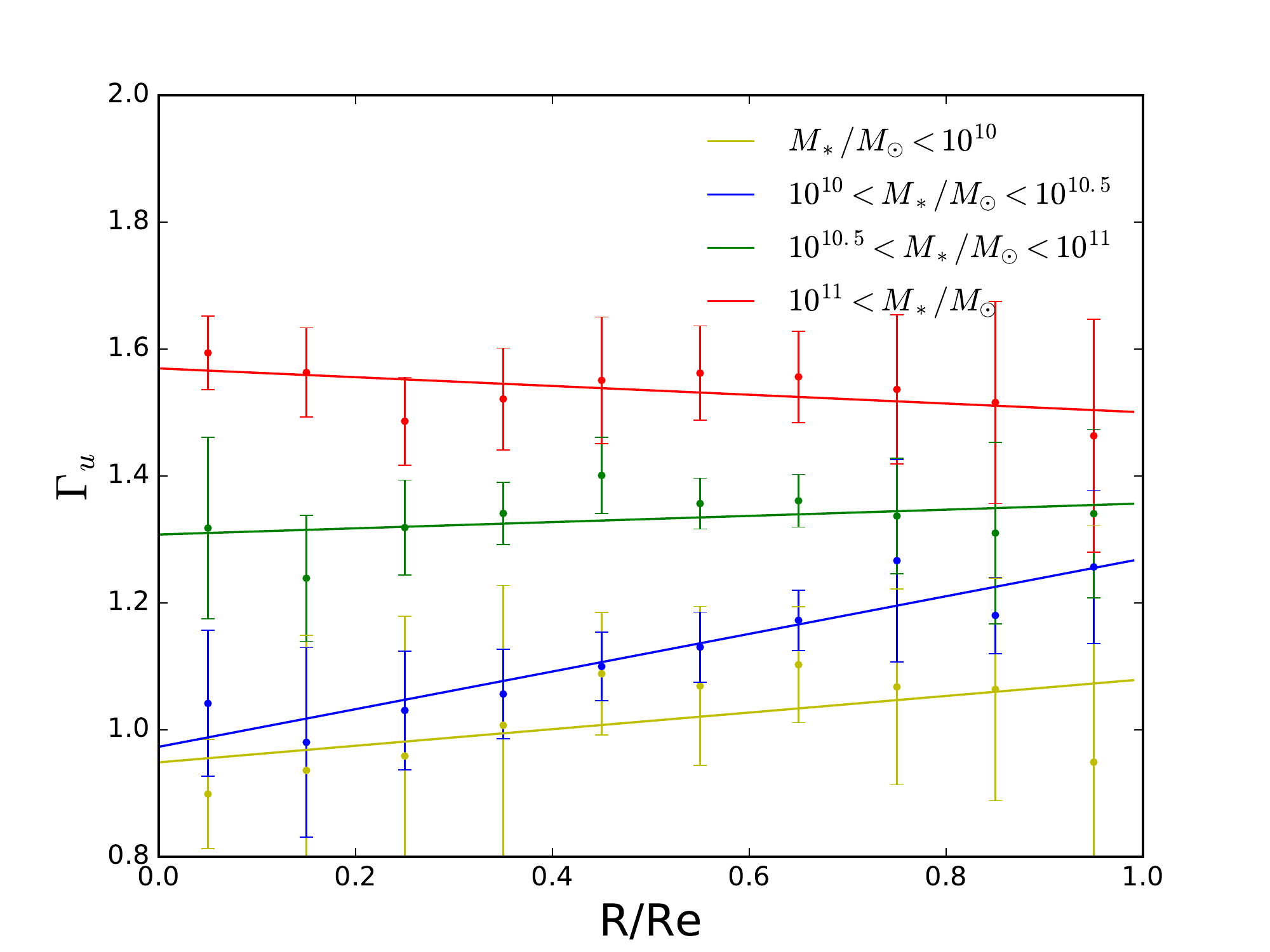}\\
\caption{The best-fit slope of the unimodel IMF, $\Gamma_u$, of MaNGA ETGs as a
function of radius. Points with different colors are results from the stacked spectra of
different stellar mass bins, as indicated in the plot. Error bars are estimated from the
jackknife method. A linear fit to these points are plotted as a reference.}
\label{radial-un}

\includegraphics[height=70mm]{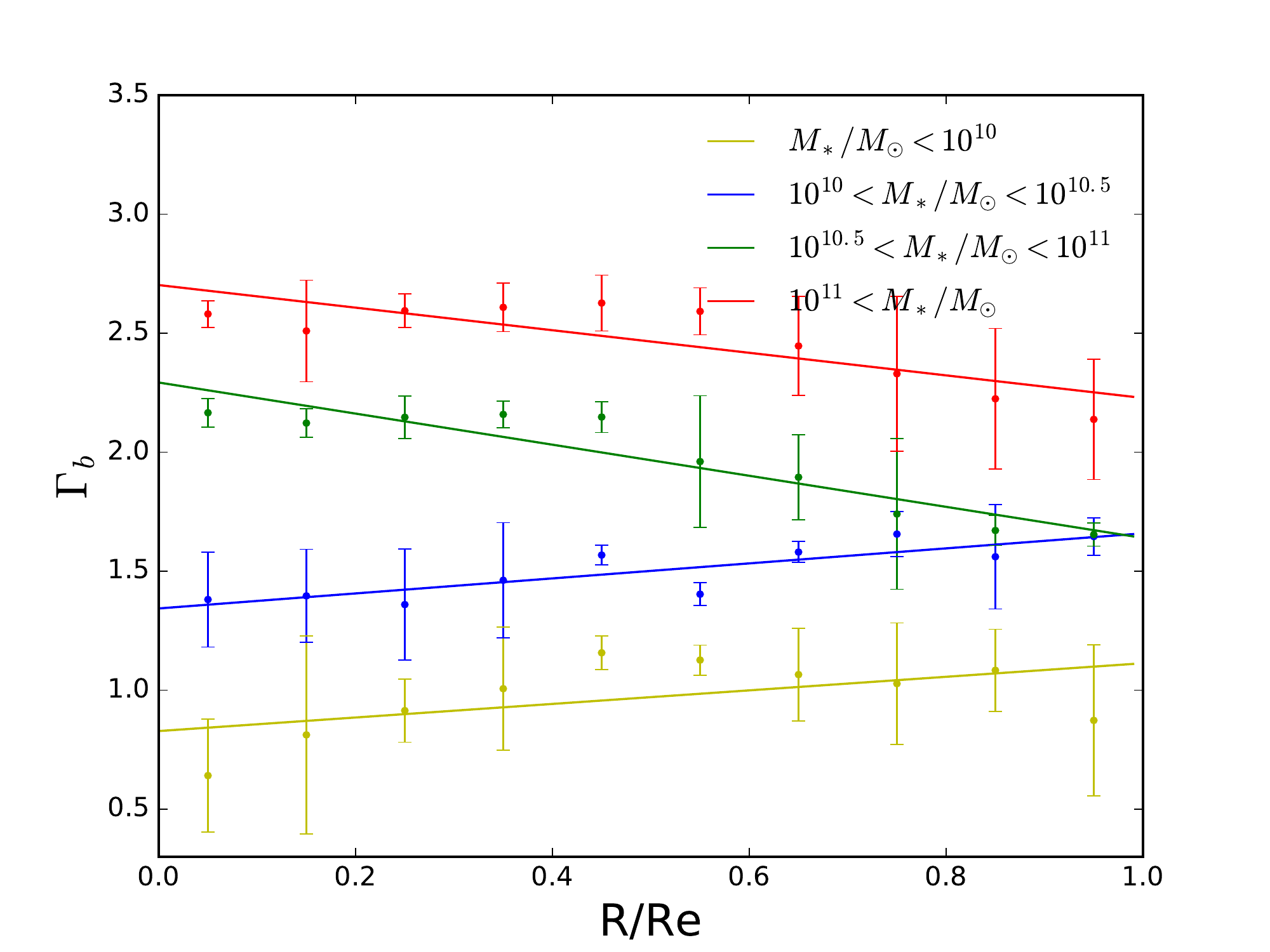}\\
\caption{The best-fit slope of the bimodel IMF, $\Gamma_b$, of
MaNGA ETGs as a function of radius. Points with different colors are results from the
stacked spectra of different stellar mass bins, as indicated in the plot. Error bars are
estimated from the jackknife method. A linear fit to these points are plotted as a
reference.}
\label{radial-bi}
\end{figure}

MaNGA IFU data allows us to study possible radial variations of the IMF slope
across individual galaxies. Here we carry out such a
study by analyzing the radially binned spectra described in \S\ref{stack}.
Results obtained from unimodel and bimodel are shown in
Figures \ref{radial-un} and \ref{radial-bi}, respectively.
To estimate the statistical error, we again use the jackknife
method to construct 20 jackknife copies.
The variance of the IMF slope values estimated from these jackknife
copies are shown in the two figures as the error bars.

Figures \ref{radial-un} and \ref{radial-bi} show a number of interesting,
albeit weak trends. First, at a given $R/R_e$ the values of
both $\Gamma_u$ and $\Gamma_b$
increase with stellar mass. This is expected from the results presented in
\S\ref{gammavsvd}, as the stellar mass of a galaxy is strongly correlated
with its velocity dispersion. Second, for massive ETGs with
$M_*>10^{10.5}M_\odot$, the IMF slope, $\Gamma_b$, tends to
decrease with increasing radius, suggesting that stars in the
inner parts of massive galaxies prefer a bottom heavier IMF than in the
outer parts. The trend in $\Gamma_u$ is rather weak, but
the error bars are larger. Third, for ETGs with $M_*<10^{10.5}M_\odot$,
there seems to be a positive trend of the IMF slope with radius,
in that the IMF is steeper in the outer part. For reference, linear
fits to the radial gradients are shown as the solid lines in the plots.
We have made tests by using stacks according to the velocity
dispersion of galaxies, and the results are found to be similar.
Negative gradients are found for high velocity dispersion
galaxies while the gradients become weaker and even positive for
galaxies of low velocity dispersion. The the error bars in the
velocity dispersion stacks are smaller than in the stellar mass
stacks, especially for high velocity dispersion stacks.
This is expected, as stellar population properties in general show
tighter correlations with galaxy velocity dispersion
than with galaxy mass. However, we only present results for stellar
mass stacks, for the convenience of comparison with other
results \citep[e.g.][]{Taniya2018}.

The stellar population gradients of ETGs have been discussed in a number of investigations
based either on broad-band photometry \citep[e.g.][]{Wu_etal2005}, long-slit spectroscopy
\citep[e.g.][]{Sanchez2007,Spolaor2009}, or IFU data \citep[e.g.][]{LHY2018,Lian_etal2018}.
The negative gradient of the IMF slope seen for massive galaxies
is consistent with the negative metallicity gradients seen in these analyses.
For low-mass ETGs, a relation between stellar mass and metallicity gradient, in that metallicity gradient is
shallower for lower masses, is also observed in some
earlier studies \citep[e.g.][]{Spolaor2009,LHY2018,Lian_etal2018}.
However, since positive metallicity gradients are observed only in the very low-mass end, the positive IMF gradient observed for low-mass ETGs is difficult to understand.
One possibility is that low-mass galaxies
have a negative gradient in dust optical depth. The positive
gradient in IMF slope can then be generated by the degeneracy
of the IMF slope with dust optical depth,
in the sense that a lower optical depth leads to to a steeper IMF
slope, as shown in Fig.\,\ref{tauvinD}. Clearly, a detailed
analysis is required to reach an unambiguous conclusion.

\subsection{Test of uncertainties}

\subsubsection{Signal to noise ratio}

\begin{figure}
\includegraphics[height=70mm]{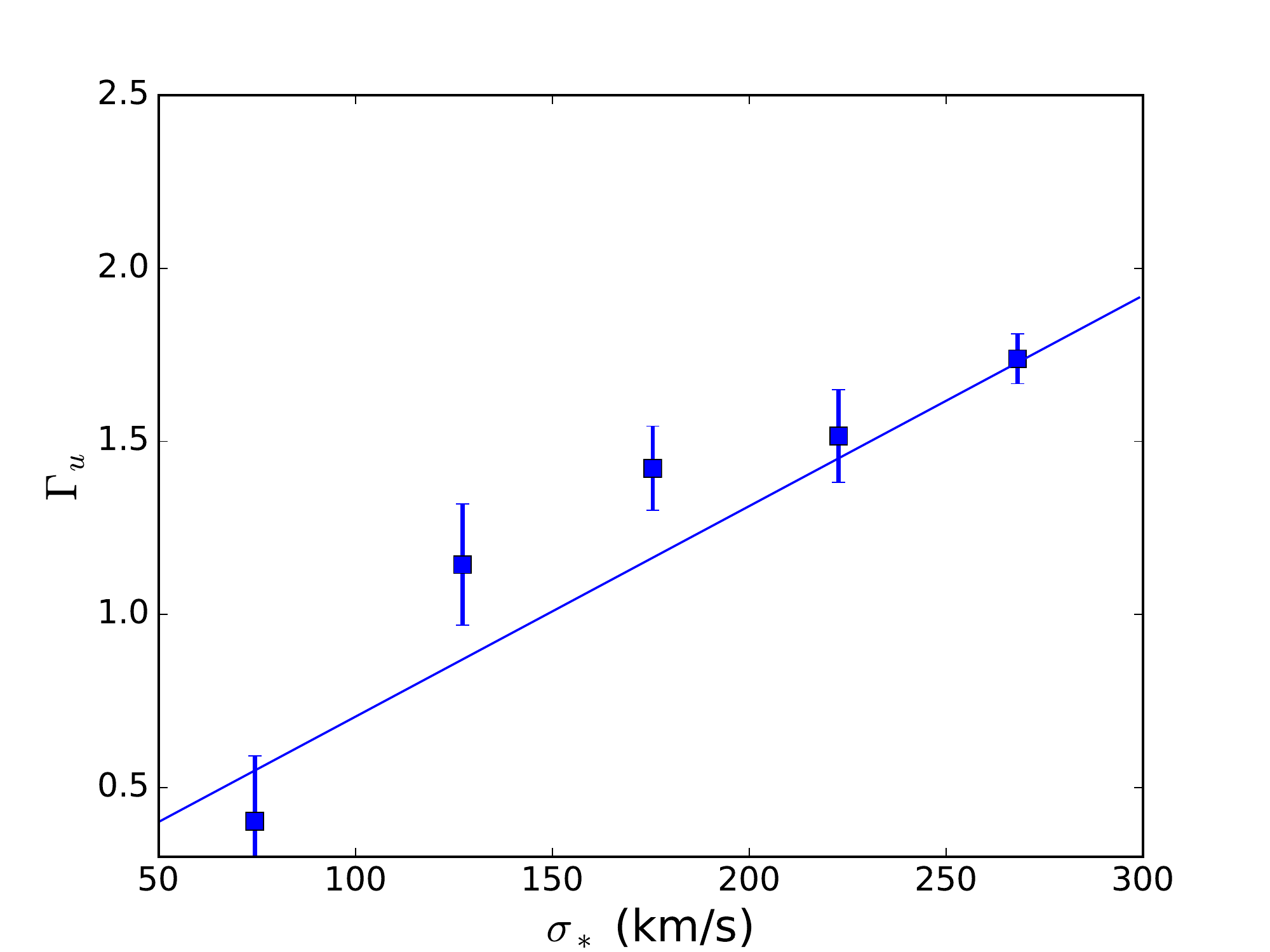}\\
\caption{The variation of IMF slope generated by random noise in different
$\sigma_*$ bins. Here the IMF slopes are inferred from a set of
artificial galaxy spectra in which random noise is added.
Blue squares are the mean slopes in the six $\sigma_*$ bins, while the
error bars show the variations generated by the noise.}
\label{bi-noise}
\end{figure}

As shown in Figs\, \ref{un}, the scatter in the IMF
slope is quite large, particularly for galaxies with low velocity
dispersion. It is, therefore, important to test whether such scatter
is produced by intrinsic variations or by noise. To do this,
we make use of the stacked spectra in different $\sigma_*$ bins
to examine how the inferred IMF slopes are affected by
noise. The stacked spectrum is regarded as noise-free
intrinsic spectra for individual galaxies in the
corresponding $\sigma_*$ bin. Random noises are then added to the
spectra according to the noise vectors of individual galaxies
to construct an artificial sample. These spectra are then fitted
to infer the IMF shapes from the artificial sample. Here
the only source of uncertainty in the inferred IMF slope
for galaxies in a given $\sigma_*$ bin is the difference in
the noise vectors and their realizations for individual galaxies
in the $\sigma_*$ bin. The results for galaxies in different $\sigma_*$ bins
are shown in Fig.\,\ref{bi-noise}. We only plot the inferred
IMF slope for unimodel IMF, as bimodel gives very similar results.
As one can see, although the global trend in
the IMF variation is well recovered from the artificial sample,
random noise contributes significantly to the scatter
(represented by the bars), especially for low-$\sigma_*$ galaxies.
This large scatter suggests that the results of the IMF slopes for
individual galaxies should be used with caution.

\subsubsection{SSP model and abundance ratio}
\label{ssec_abundence}

\begin{figure}
\includegraphics[height=70mm]{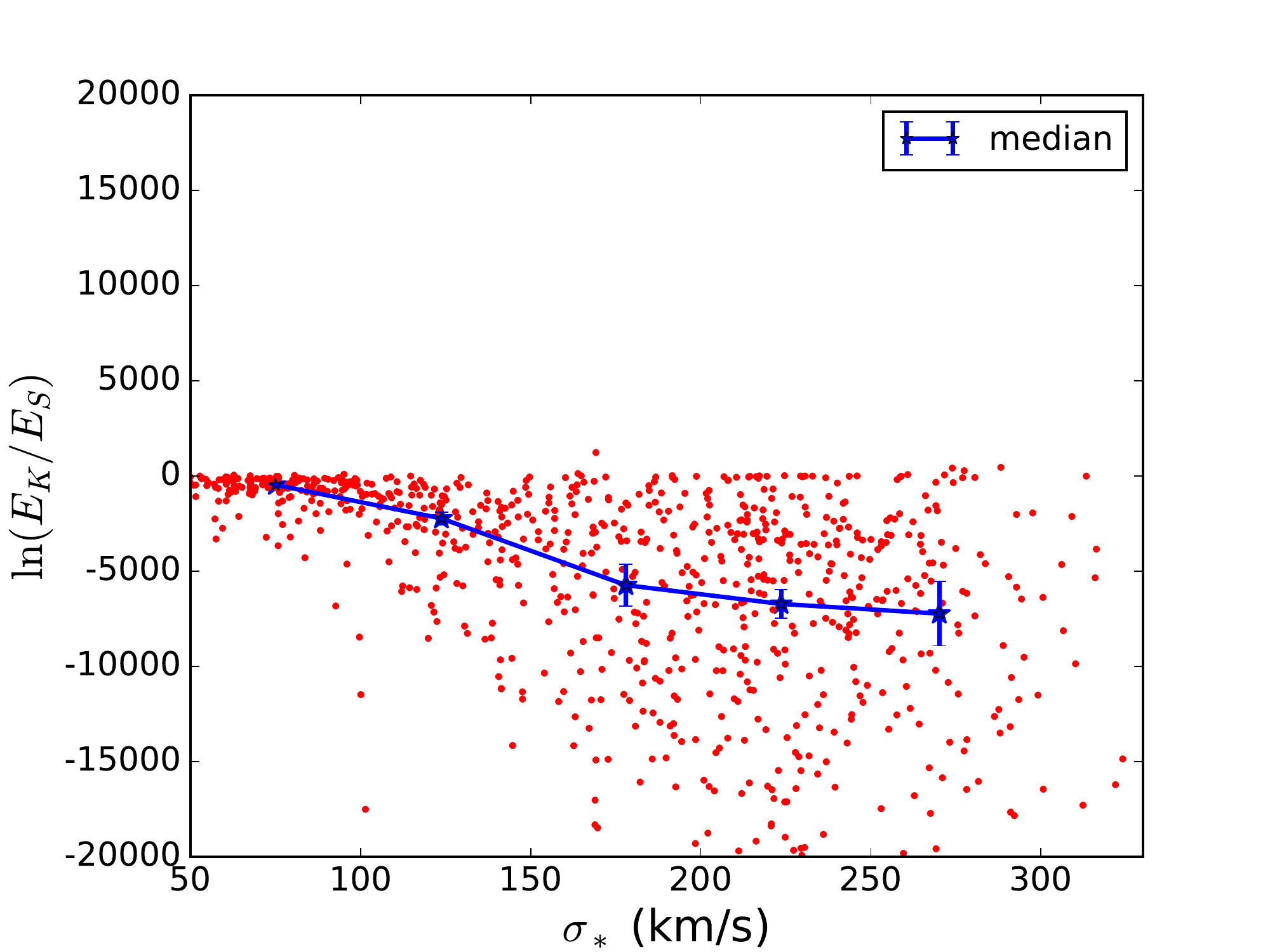}\\\
\caption{The evidence ratio between Kroupa and Salpeter IMFs plotted
against galaxy velocity dispersion, as derived from the C18
model. Each red dot stands for the result of a MaNGA
ETG. Blue stars are the median values in five $\sigma_*$ bins
and connected by a blue line. Error bars are  obtained from the
jackknife method.}
\label{evratio-c18}
\end{figure}

\begin{figure}
\includegraphics[height=70mm]{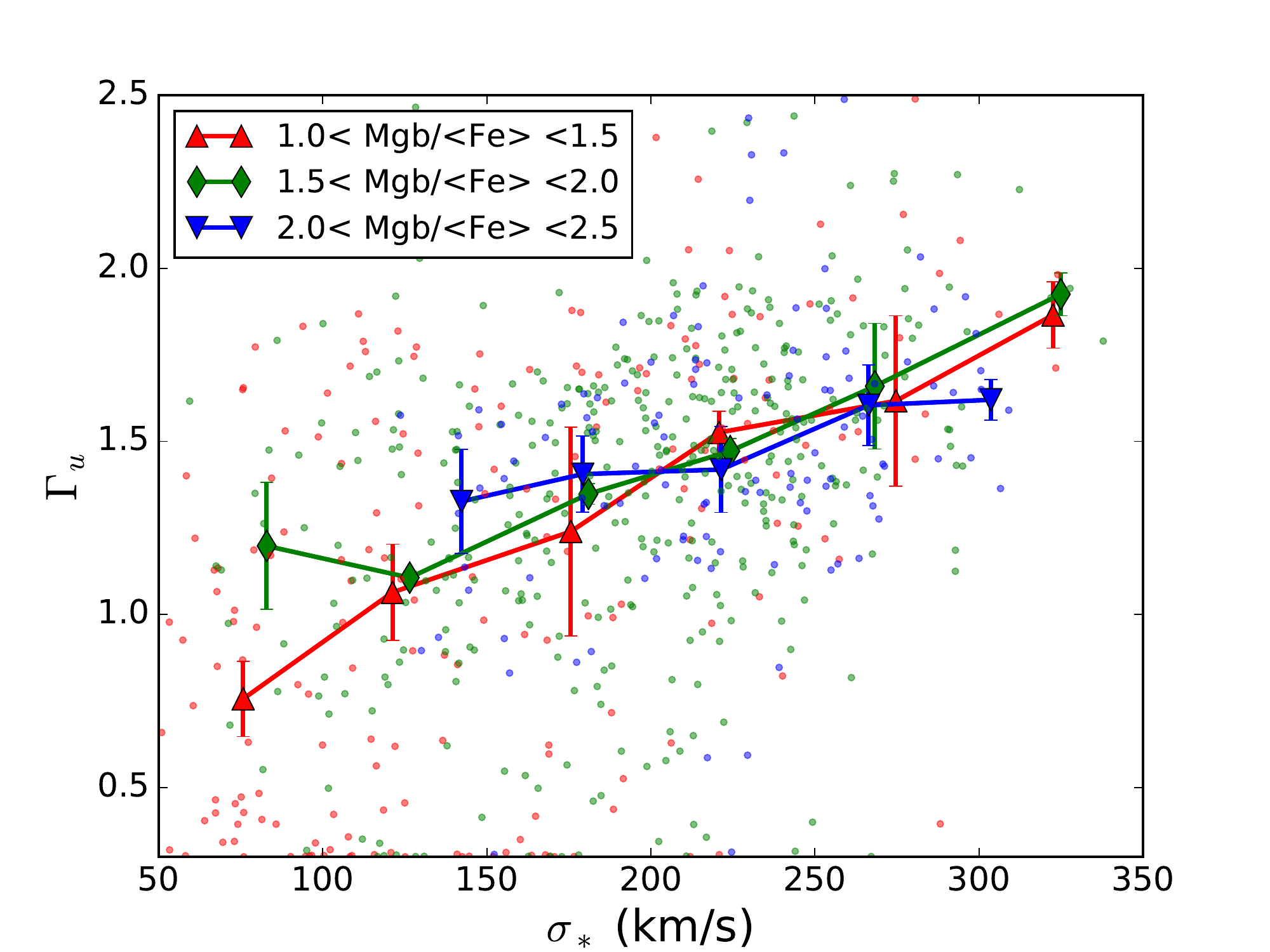}\\
\caption{The variation of $\Gamma_u$ for galaxies in different
$\rm Mgb/\langle Fe\rangle$ bins.
The red, green, blue dots stand for
galaxies in different $\rm Mgb/\langle Fe\rangle$
bins, as indicated.
Big symbols connected by thick lines are results from the stacked
spectra in the corresponding
$\rm Mgb/\langle Fe\rangle$ bins.
Error bars are estimated from the jackknife method.
}\label{bi-alpha}
\end{figure}

\begin{figure}
\includegraphics[height=70mm]{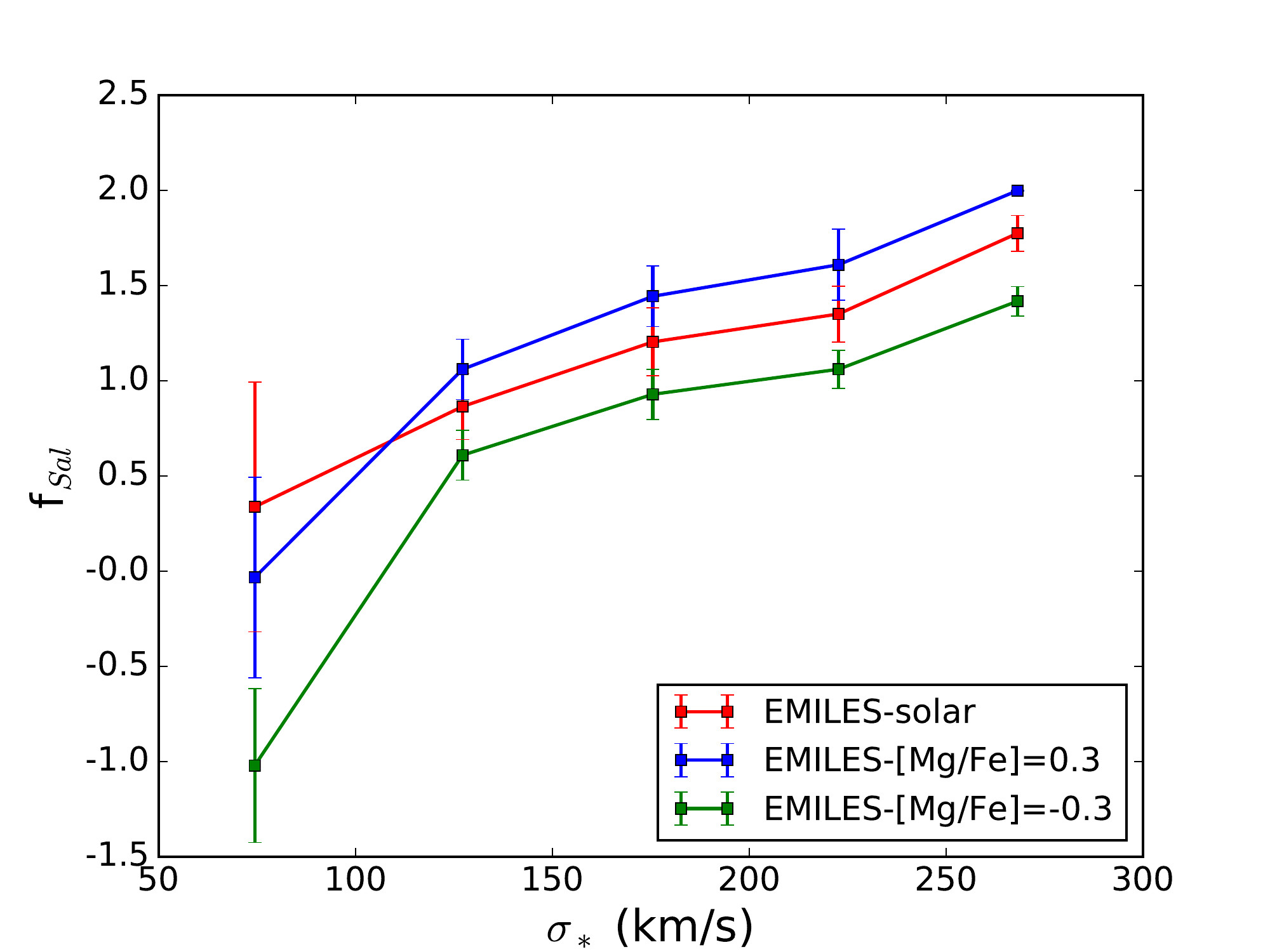}\\
\caption{The relative contribution of the population with Salpeter IMF, $f_{\rm Sal}$, inferred from the E-Miles models with
three different [Mg/Fe], as indicated. Error bars are estimated from
the jackknife method.
  }\label{fsal-mgb}
\end{figure}

When model uncertainties are ignored, Stellar population parameters
recovered from full-spectrum fitting are in general stable
if the spectrum has sufficiently high SNR \citep{Ge2018}.
However, the SSP model adopted in the fitting can affect the fitting
results significantly. The SSP models can differ in a variety of ways,
such as stellar templates and isochrones, and even in the
method to populate stars in parameter space.
In addition, public versions of different SSP models generally
have different parameterizations of the IMF for different
purposes. All these can potentially affect the inferences of
model parameters from the observed spectra. Efforts have been made
to ensure the consistency of the IMF inferred from different
stellar population models. For example, \cite{Spiniello2015} compare two
sets of widely-used SSP models to demonstrate the
non-universality of the IMF in the low-mass end, and
conclude that the non-universality of the IMF is robust
against the choice of SSP model. For the MaNGA data,
\cite{Taniya2018} use two different SSP models,
VCJ \citep{Villaume2017} and M11-MARCS \citep{Maraston2011,Gustafsson2008},
to fit IMF-sensitive line indexes
in the near infrared spectral region, and obtain
qualitatively consistent results. All these indicate that
different SSP models agree with each other in general,
while discrepancy may exists between models in detail.

Since these results are all based on absorption lines, it remains
unclear whether or not our results based on full spectrum fitting can
be affected significantly by our adopted SSP model.
To test this, we carry out an analysis by using the SSP
model presented in \cite{Conroy2018} (referred to as C18 model)
instead of the E-MILES model. Different from E-MILES model, C18
is constructed from the MILES and Extended IRTF libraries
and has a number of free parameters to describe
the variations in element abundance patterns but
only provides a limited choices for the IMF.
To make a fair comparison with Fig.\,\ref{evratio},
we restrict our analysis to the base C18 model, which assumes
solar abundance pattern and Salpeter and Kroupa IMFs, neglecting all the
free parameters that account for element abundance variations.
These SSP models are then used to fit the stacked spectra,
as was done in Fig.\,\ref{evratio}, and the resulting evidence ratio
between the Kroupa and Salpeter IMF models, $\ln(E_K/E_S)$,
versus $\sigma_*$ is shown in Fig.\, \ref{evratio-c18}. Comparing the
result with that shown in Fig.\,\ref{evratio}, we see that
the trend that the Salpeter IMF is preferred by high-mass
ETGs are not affected significantly by the change of the
SSP model.

Another potential uncertainty comes from possible variations of
the element abundance pattern from galaxy to galaxy.
It is easy to understand that the inference of the IMF from absorption
line analysis is sensitive to the abundances of individual
elements, especially those whose absorption lines are used
(e.g. [Na/Fe]). In the full spectrum fitting adopted here,
the problem may be alleviated, as the influence of the variation
of individual element abundances is generally limited to specific
wavelength ranges. However, a systematic variation of a group of
elements, such as the $\alpha$ elements, may still cause significant
difference in the continuum, thereby affecting our inferences
for the IMF shape. Moreover, elliptical galaxies are known to
be $\alpha$-enhanced \citep[e.g.][]{Trager2000},
while the E-MILES templates adopted here
are constructed with the solar abundance pattern. It is
therefore necessary to test whether our inferences are
affected significantly by the change of the abundance ratio.

The lick index ratio ${\rm Mgb/\langle Fe\rangle} \equiv \rm 2Mgb/(Fe5270+Fe5335)$
has been widely used as a good indicator for the $\alpha$-abundance,
especially for old populations (ages above $\sim 8$ Gyr,\citealt{Thomas2003}).
Fig.\,\ref{bi-alpha} shows results for galaxies divided
into three bins according to their ${\rm Mgb/\langle Fe\rangle}$.
It is clear from the plot that the trends of the IMF slope with velocity dispersion
are similar without strong dependence on ${\rm Mgb/\langle Fe\rangle}$,
indicating that our results are insensitive to the
assumed element abundance pattern. We have also examined the
same correlations as shown in Fig.\,\ref{bi-alpha} but using
$\alpha$-abundance estimates of \citet{Zheng_etal2019} obtained
from SPS models. Although not shown here, the trends obtained are
very similar to those shown in Fig.\,\ref{bi-alpha}.

 A more direct way to check the impact of abundance ratio
is to use SSP models with varying abundance ratios.
Unfortunately, E-MILES does not provide templates
with different [$\alpha$/Fe] or other abundance ratios.
As an alternative test, we use the response functions from
\cite{Conroy2018} to construct models with different abundance
patterns. The response functions are ratios between
the SSPs generated with stellar templates with
varied abundance for a single element and those generated with
the standard abundance patterns. SSP models with different abundance
patterns can then be generated by multiplying those response functions
with the original model. These functions cover a wide range in metallicity
($-1.5<{\rm [Fe/H]}<0.3$), and are appropriate for old stellar
populations with ages $>1{\rm Gyr}$.
Among all the element abundance variations,
that of [Mg/Fe] is the easiest to measure and has significant
impact on the spectral shape. As a preliminary test, we use the response
function of Mg/Fe from \cite{Conroy2018}, which is given for
[Mg/Fe]=$\pm0.3$, to generate two perturbed E-MILES models.
As the response function is available only for a limited number of
IMF models, we focus on the widely used Kroupa and Salpeter IMFs.
To quantify the IMF variation, we model a given
spectrum with $F=(f_{\rm Sal})\times F_{{\rm Sal}}+(1-f_{\rm Sal})\times F_{\rm Krp}$,
where $F_{\rm Sal}$ and $F_{\rm Krp}$ are spectral fluxes predicted
with a given set of parameters but from SSPs assuming
Salpeter and Kroupa IMFs, respectively. The parameter $f_{\rm Sal}$,
which can be interpreted as the relative contribution of the
population with a Salpeter IMF, then takes the place of the IMF
slope used in previous sections to characterize the IMF
shape. As $f_{\rm Sal}$ is used to form a linear combination of
the two IMFs, the value of $f_{\rm Sal}$ is allowed to be bigger
than one and smaller than zero.

The three SSP models, namely the original E-MILES model, and
the two perturbed models with ${\rm [Mg/Fe]}=\pm 0.3$,
are then used to fit the the stacked spectra, and
the results are shown in Fig.\,\ref{fsal-mgb}. It can be seen
that $f_{\rm Sal}$ is positively correlated with $\sigma_*$
for all the three assumptions of ${\rm [Mg/Fe]}$, indicating again
a bottom-heavier IMF for more massive ETGs. Comparing results
of different ${\rm [Mg/Fe]}$, we see that the value of $f_{\rm Sal}$
increases systematically with the abundance ratio ${\rm [Mg/Fe]}$.
However, the overall trend of $f_{\rm Sal}$ with $\sigma_*$
remains unchanged by the change in ${\rm [Mg/Fe]}$.
Since ${\rm [Mg/Fe]}$ is not expected to be anti-correlated with
$\sigma_*$ \citep[see][]{Zheng_etal2019} for the MaNGA galaxies,
the trends seen in Fig. \ref{un} are robust
against [Mg/Fe] variations.

This test can be applied to all the available element abundance variations separately.
Although not shown in the paper, we have done the same tests for abundance ratios
like [Na/Fe] and [Ca/Fe], and found that the impact of these variations are much
smaller. One can in principle include all abundance ratios simultaneously under the
assumption that the responses from different elements are uncorrelated. In fact,
such an approach has already been adopted in some earlier IMF investigations
\citep[e.g.][]{Conroyb2012,Villaume2017imf}. However, as our Bayesian analysis
needs to estimate multi-dimensional integrations to obtain the global evidence for a
large sample of ETGs, it is extremely time consuming to include all possible abundance
variations in the fitting. Furthermore, although we have limited to Salpeter and
Kruopa IMFs to avoid differences in the IMF parameterization, the E-MILES and C18
models may have other differences that have not been explored,
and it is unclear whether the response functions taken from C18 are accurate
for E-MILES. As our tests show that our conclusions are not affected
significantly by element abundance variations, we will not explore this
issue in more detail.

\subsubsection{Star formation history}

Our inferences presented above assume that the star formation
histories (SFH) of individual galaxies can be described by the
$\Gamma$ function model, equation (\ref{gamma-sfh}). This may be
a good approximation for massive elliptical galaxies, where
stellar populations are old and more or less uniform.
For low-mass galaxies, however, the star formation histories
may be more complex. To test the effects of the assumed star formation
histories on our inferences of IMF slopes,  we have run
a set of spectral fitting by assuming the widely used
$\tau$ model, referred to as $\tau$-SFH, which can be written as
\begin{equation}\label{SFH_tau}
\Psi(t)=\frac{1}{\tau}\frac{e^{-t/\tau}}{1-e^{-\Delta t/\tau}}
\end{equation}
where $\tau$ is the star formation time scale and $\Delta t$ is
the time span over which star formation occurs.
We find that the results obtained by using the
$\tau$-SFH are all consistent with the results obtained from
the $\Gamma$-SFH, indicating that our results are
not affected significantly by the change of the SFH model.

In addition to the two continuous star formation histories, we have
also tested whether or not an additional burst population is needed by the data.
By comparing the evidence ratio between models of the $\Gamma$-SFH
with and without the additional burst, we find that the majority of
the galaxies do not show a preference for an additional stellar population.
In summary, the SFH model adopted in our analysis is sufficient for
our purpose. A more detailed modeling of the SFH may improve the inferences
for some galaxies, but it will not affect our main conclusions.

%
\section{Comparison with earlier results}
\label{sec_comparisons}

Our analysis is based on the MaNGA data and fitting to the full
spectra. It is complementary to earlier studies based on different methods
and datasets. In this section, we compare our results with those obtained
before.

\subsection{The IMF slope}
\label{sec_comparisons_slope}
Our analysis reveals a positive correlation between the IMF slope, which
describes the relative ratio between low-mass and high mass stars,
and the central stellar
velocity dispersion for ETGs over a large range of velocity dispersion,
$\sigma_*\approx 50-350$ \kms.  This excess of low mass stars in high mass ETGs
is also supported by Bayesian
model selections between a Milky Way like IMF and a bottom-heavy IMF.
This correlation has been found earlier in a number of investigations
\citep[e.g.][]{Thomas2011,Dutton2012,Cappellari2012,Conroyb2012,Spiniello2012,Ferreras2013,Barbera2013,
Posacki2015,Lyubenova2016,lhy2017,Villaume2017imf,Taniya2018}.
Among these, \cite{Ferreras2013} used a SSP model family
(the MIUSCAT SSP models, a progenitor of E-MILES) similar to
what is adopted here, to analyze the stacked spectra from SDSS
for galaxies with $\sigma_*>150 $ \kms.
Their results,  obtained from TiO1, TiO2 and Na8190 absorption
features versus ours based on full spectrum fitting,
are in quantitative agreement, suggesting consistencies
in the two approaches. However, the single fiber spectroscopy
did not allow them to investigate possible IMF variations within
individual galaxies.

\cite{Barbera2013} also used the MIUSCAT SSP models to study the IMF variation,
taking into account possible non-solar abundance. Their spectral analyses
did not seem to be able to discriminate a single power law (unimodal) IMF
from a low-mass ($<0.5 M_{\odot}$) tapered (bimodal) IMF, in contrast to
our Bayesian analysis that clearly favours unimodal IMF, especially for
high mass ETGs. However, if dynamical constraints are taken into account,
as discussed in \cite{Barbera2013}, unimodal IMF models can lead to unrealistic
mass-to-light ratios that are too large to be consistent with dynamical measurements,
while bimodal IMF models well meet the dynamical constrains.
Similar results were reported in \cite{Lyubenova2016} by comparing dynamical
modeling and stellar population results obtained from applying E-MILES models
to CALIFA data. The tension between stellar population and dynamical results is
likely due to issues in IMF parameterizations. In fact, as shown in \cite{Conroyb2012},
a three component power-law IMF with slope varying at the low mass end
($0.1 M_{\odot}< M < 0.5M_{\odot}$ and $0.5M_{\odot}< M <1.M_{\odot}$)
but fixed at the high mass end ($>1M_{\odot}$), seems to be able to fit
the absorption features in ETGs and to satisfy dynamical constraints
simultaneously. Unfortunately, the bimodal IMF adopted here
does not really have the flexibility at the low-mass end.
We suspect that this is the reason why the uimodal IMF is preferred
in our analysis. Because of the single IMF slope over the entire mass range,
the uimodal IMF may fail to reproduce the mass-to-light ratios of galaxies.
Unfortunately, the lack of a more flexible IMF parameterization in
E-MILES model prevent us from investigating this issue further.

\cite{Taniya2018} used a MaNGA data set similar
to the one used here to investigate variations of the IMF.
Their analysis was based on the absorption lines, NaI and FeH,
and the SSP templates of VCJ \citep{Villaume2017} and M11-MARCS
\citep{Maraston2011,Gustafsson2008}, different from the
E-MILES templates adopted here. The general trend they found,
that the type of IMF seems to change from low mass to high mass
ETGs, is similar to what is found here, indicating that the
results are not sensitive to the SSP models adopted.
However, the radial gradients in the IMF slope they found
are quite different from what we obtain here, as to be
described in more detail below (\S\ref{ra}).

In addition to the correlation between the IMF slope and
$\sigma_*$, other types of correlation, especially  the IMF-metallicity correlation, have also been suggested.
For example, studies of Wolf-Rayet stars indicate a correlation between gas phase
metallicity and IMF \citep{Zhang2007}, and direct star counts in nearby resolved galaxies
\citep{Geha2013} give a similar correlation between IMF and stellar metallicity.
Such a correlation is also revealed in the study of 24 CALIFA galaxies by \citet{Navarroa2015}, who found that the
correlation of the IMF shape, $\Gamma_b$, with galaxy metallicity is
the strongest among the correlations with a number of other galaxy properties,
such as velocity dispersion. Similar conclusion was reached
by \citet{Taniya2018} using radially stacked spectra of MaNGA
galaxies. \citet{Villaume2017imf} studied a parameter,
$\alpha_{\rm IMF}$,  which is the ratio of the stellar mass-to-light ratio,
$(M/L)_*$, obtained from fitting with varying IMF, to
the value of $(M/L)_*$ obtained with a MW IMF, and its correlation
with [Fe/H], [Mg/Fe] and $\sigma_*$, for a sample that includes
ETG galaxies from \citet{vanDokkum2017}, three M31 globular clusters, one
ultra-compact dwarf, and M32. They found that the IMF variation
is most tightly correlated with [Fe/H]. Our results confirm
these results with a much larger sample covering a larger dynamic range.
In particular, our Bayesian approach allows us to examine in detail the
covariance and degeneracy in the various correlations seen in the data.
\cite{Villaume2017} also found a globular cluster that
deviates significantly from the global trend, and suggested that
metallicity is not the sole driver of IMF variation.

From a theoretical point of view, the thermal properties of star-forming clouds
are expected to have important influences on how they fragment into stars,
which can, in turn, affect the stellar IMF \citep{Larson2005}. Since the
abundances of atoms, molecules, and dust grains are responsible for the cooling,
the overall metallicity is expected to play an important role in shaping the IMF.
However, only limited attempts have been made in quantifying the effect of metallicity
on the thermal properties of collapsing clouds \citep[e.g.][]{low1976, Omukai2000}.
As our results provide strong evidence that metallicity is the dominant driver of
IMF variations, which is consistent with other studies mentioned above, we expect
to see further studies of the thermal properties of star-forming clouds in
different circumstances and how the stellar IMF can be affected by metallicity.

 Finally, a completely different approach, based on dynamical modeling
of MaNGA galaxies, has been adopted by \cite{lhy2017} to constrain
potential IMF variations. The dynamical modeling provided
constraints on the stellar mass to light ratio of galaxies
as a function of galaxy velocity dispersion, which were then
compared to the expected mass to light ratios from different
IMFs. The trend they found is qualitatively the same as
found here, in that galaxies with higher velocity dispersion
tend to prefer a bottom-heavier IMF. However, as the kinematic
results can only provide an overall constraint on the stellar mass
to light ratio, they may be subjected to a different set
of uncertainties, such as the uncertainty in the modeling
of dark matter distribution, and the degeneracy between
IMF and star formation history in the mass-to-light ratio.
Because of this, a quantitative comparison between their
results and ours is difficult to make.

\subsection{Radial dependence}
\label{ra}

Thanks to the spatially resolved IFU spectra provided by
MaNGA, we are able to derive radial gradients from a set of stacked
spectra. Our results show a difference in the radial gradient,
from a negative gradient to a positive gradient in
the IMF slope, between massive ETGs with $M_*$ larger
or smaller than $10^{10.5}M_{\odot}$.
A number of earlier investigations have reported the presence of IMF
gradients indicating dwarf-richer stellar populations towards the centers
of massive ETGs \citep[e.g.][]{Navarrob2015,Barbera2017,vanDokkum2017,Vaughan2018b,Sarzi2018,Taniya2018},
with the exceptions in some other investigations \citep{Vaughan2018,Zieleniewski2015,Zieleniewski2017,McConnell2016,Alton2017}.

Many of the earlier investigations of the IMF gradient are based
on small samples, as IFU observations are relatively rare.
\cite{Navarrob2015}, for example, studied the radial gradients of
the IMF slope, $\Gamma_b$, for three early-type galaxies with central
velocity dispersion
$\sigma_*\approx 300$, $280$, and $100$ \kms, respectively,
and found strong enhancements of low mass stars in the central
parts of the two large galaxies, but a flat or slightly opposite
trend for the galaxy with the lowest
$\sigma_*$. Their results are based on spectral indices including
TiO1, TiO2, NaD and NaI8190. Although the last two indexes may be
affected by sky emission
and telluric absorption, the trends they found
for the three galaxies are in good agreement with our results.
A similar analyses, carried out by \cite{Barbera2017} using spectra
indices in redder bands, showed that IMF varies with radius
in two nearby massive ETGs with $\sigma_*\approx 300$ \kms.
\cite{vanDokkum2017} studied six ETGs with $\sigma_*$ ranging
from 163 to 340 \kms, using IMF-sensitive absorption lines
instead of spectral indexes. After accounting for element abundance
gradients, they were able to find strong enhancements of low mass
stars in the central parts of all the six galaxies.

Compared to these case studies using small number of galaxies,
large IFU surveys are complementary and particularly suitable for
statistical analyses. However, observational samples have
been quite limited up to now. \cite{Navarroa2015} analyzed a sample of 24 ETGs
with $\sigma_*$ ranging from 160 to 310 \kms and a mean stellar mass
$10^{11.54}M_{\odot}$, drawn from the CALIFA \citep{CALIFA2012} IFU survey.
They found that the local IMF within a
galaxy is tightly correlated with the local metallicity,
being bottom-heavier for metal-richer populations.
As ETGs generally have negative gradients in metallicity, their results
imply a IMF gradient in massive galaxies in the same way as we find.

As mentioned above, \cite{Taniya2018} used a set of MaNGA galaxies
similar to ours to study radial gradients of IMF by modeling absorption
line indices. Their analysis based on NaI index and the M11-MARCS
stellar population model revealed a significant negative gradient
in massive galaxies, while a much weaker gradient is found for
low-mass galaxies. This is in good agreement with ours, suggesting
a difference in IMF gradient between high and low mass galaxies.
However, their analysis of the same index but using the VCJ stellar
population model shows a significant negative gradient also
for low-mass galaxies. This suggests that the evolution of low-mass
galaxies may be more complex, and thus is more challenging to model.
More surprisingly, their analysis of another IMF-sensitive index,
FeH at around 9900\AA, using either of the two SSP models,
revealed strong {\it positive} radial gradients in the IMF slope,
in conflict with their results based on the NaI index and the results
we obtain. This may arise from the difficulty in modelling reliably the FeH
index. Our fitting does not cover that spectral region, and thus
would not suffer from this inconsistency.

While many previous studies have found evidences
for the presence of negative IMF gradients, a few of the
investigations questioned the significance of such gradients.
These investigations generally
did find radial gradients in absorption features. However, the
gradients can be interpreted as those in element abundance.
For example, \cite{Alton2017} studied eight early-type galaxies
with $\sigma_*\approx250$ observed in $0.8-1.35{\rm \mu m}$, and
concluded that the measured radial changes in absorption feature
strengths can be explained entirely by abundance gradients.
In addition, they found that the fractional contribution of dwarf
stars to the $J$-band total flux, which is believed to be closely
related to the IMF, also provides no clear evidence for
radial variations of the IMF. However, the validity of their inference
of low-mass star fraction can be limited by the wavelength
coverage, as it can be affected by uncertainties in age and
metallicity. Additional information, such as spectral indexes or
spectra covering shorter wavelengths, may be needed to break the
degeneracy \citep[e.g.][]{Vaughan2018}.

In conclusion, the radial dependence of the IMF we find for ETGs
is consistent with some of the earlier results. The fact that
there is significant variation in the IMF shape among individual
galaxies of similar $\sigma_*$ and that the radial IMF gradient
on average depends on galaxy mass suggests that
some of the discrepancies seen in earlier results may be
due to such variation and dependence.

%
\section{Summary and discussion}
\label{summary}

We use a Bayesian inference code to analyze the spectra of a sample of
ETGs selected from the SDSS-IV MaNGA IFU survey and to study
possible variations in their IMF. High SNR spectra of individual
galaxies are obtained by stacking the spatially resolved spectra within
$1.0 R_e$,  as well as by radially stacked spectra for a sample of galaxies
in given stellar mass bins. Our analysis is based on full
spectrum fitting, making use of the MaNGA spectra
from 3400 \AA \ to 8900 \AA, while masking the problematic region
(6800 \AA \ to 8100 \AA). The state-of-the-art E-MILES synthetic
stellar population models, with two types of IMF parameterizations,
unimodel and bimodel, are used to infer the IMF slope from
the spectra, while the widely used Chabrier and Salpeter IMFs are
tested with Bayesian model selections. Our main results can be summarized
as follows:

\begin{itemize}
\item
The IMF is not universal among ETGs, as indicated by the evidence ratio
between different types of IMF. A clear trend of the evidence ratio
between Chabrier and Salpeter IMFs with galaxy velocity dispersion,
$\sigma_*$, shows that the Salpeter IMF becomes more
favored over the Chabrier IMF for higher mass ETGs.  In addition,
the unimodel model family, which is described by a single power
law, is also favored by massive ETGs over a model family
in which the IMF is assumed to flatten in the low-mass end.
All these suggest that Galactic-like IMF may be invalid for
high-mass ETGs.
\item
The IMF slope inferred from our full spectra fitting follows a clear trend with
central velocity dispersion $\sigma_*$, in the sense that galaxies
of higher $\sigma_*$ on average have IMFs that are more bottom-heavy.
However, this correlation is driven by the dependence of IMF shape
on metallicity, as galaxies of similar metallicities have only weak
dependence on $\sigma_*$.
\item
The degeneracy between metallicity, age and dust extinction
of a galaxy can induce systematic trends in the IMF
slope with stellar age and dust optical depth. The dependence of the
IMF slope on these parameters is through a combination,
$D$ that is perpendicular to the degeneracy direction.
\item
Using age and metallicity estimates in which the degeneracy
between them is reduced, and their scaling relations with galaxy
velocity dispersion, we found the the IMF variations
can be explained entirely by metallicity variations,
but not by age variations.
\item
We found a weak negative gradient in the IMF slope in individual galaxies,
in the sense that the IMF in the center of a massive ETG is,
on average, more bottom-heavy than in the outer part,
while a weak opposite tend is found for low-mass ETGs. This
dependence of IMF radial gradient on galaxy mass suggests
that part of the discrepancies seen in earlier results may be due
to such dependence
\item
We tested all the trends found in the paper with both  IMF
parameterizations, unimodel and bimodel, and found no significant
differences. In addition, we also tested the robustness of our
results against a number of uncertainties, such as the variation
in spectral signal-to-noise ratio, stellar population models with
varying abundance ratios, and the modelling of star formation
histories.
\end{itemize}

 Our investigation is one of the first attempts to use full spectrum
fitting to constrain the IMF shape of galaxies. With the help of Bayesian
analysis, we are able to compare different IMF model families, as well
as to examine how the inferred IMF slopes are affected by the degeneracy
with other model parameters. This is a novel approach to the problem,
and here we discuss further some of the potentials and pitfalls of the
method.

 Compared to absorbtion line analysis, the full spectrum fitting method
makes use of more information contained in the spectra. This allows one
to determine the IMF shape together with other properties, such as stellar
age, metallicities, dust attenuations, and to examine their covariance.
In fact, some earlier studies \citep[e.g.][]{Ferreras2013} have
used the optical spectra to measure auxiliary galaxies properties
in addition to the IMF. The Bayesian approach adopted here is a
statistically more rigorous way to make self-consistent inferences
from the full spectra. As shown in the paper, the IMF slope and its
covariance with stellar age, metallicity and dust extinction
can be quantified in the Bayesian approach. In particular,
the covariance is found to be only modest, and does not hinder
the inferences of the IMF shape.

 An issue in absorption line analysis is that it can
be affected significantly by the abundance variations of
certain elements. Detailed theoretical modelling of a small
number of absorbtion features is possible
\cite[e.g.][]{Conroya2012,Conroy2018}, and may help to take
into account such abundance variations.  In full spectrum fitting,
however, theoretical templates are not commonly used, as
it is hard to model the large number of spectral features in
the whole spectra. Currently, the lack of observed stellar templates
of varying element abundance patterns weakens the abilities of
full spectrum fitting in inferring the abundance of elements that
might be related to the IMF. However, the influence of element
abundance variations is generally confined to certain
spectral regions, and their effects on IMF inferences
based on full spectrum fitting are reduced, as demonstrated
by our tests. Nevertheless, it should be emphasized that a
detailed analysis of abundance ratios is needed in order to
make accurate inferences of the IMF shape, and to quantify
the covariance between IMF shape and abundance ratio.

Since IMF variation has the greatest impact in the near infrared
parts of galaxy spectra and the effect is in general small, modeling
the effect is a challenge to the spectral population synthesis model.
Most of the stellar
templates currently available (e.g. MILES) have proper
coverage only in optical, and need to be matched with
other stellar libraries to construct SSPs with appropriate
spectral coverage. Such a match can be made only when
fluxes in different wavelength ranges are calibrated properly.
Some investigators have chosen to split the spectra into pieces and
fit the local continuum \citep[e.g.][]{vanDokkum2017,Vaughan2018},
while we have used a carefully designed mask to get rid
of problematic spectral portions. Both methods may have
their own uncertainties. The fact that these approaches produce
consistent results indicates that the inferences of IMF
variations are robust against such uncertainties.
In the future, when stellar templates with higher resolution
and wider spectral coverage, such as the MaSTAR \citep{Yan2017},
become available, our method should be able to provide
an even more powerful tool to investigate IMF variations.

\section*{Acknowledgements}
This work is supported by the National Key R\&D Program of China
(grant Nos. 2018YFA0404503, 2018YFA0404502, 2018YFA0404501), the National Key Basic
Research Program of China (grant No. 2015CB857004), and the National
Science Foundation of China (grant Nos. 11233005, 11621303, 11522324,
11421303, 11503065, 11673015, 11733004, 11320101002, 11821303, 11333003, 11390372 and 11761131004 ). HJM acknowledges
the support from NSF AST-1517528.

Funding for the Sloan Digital Sky Survey IV has been provided by the Alfred P.
Sloan Foundation, the U.S. Department of Energy Office of Science, and the Participating Institutions.
SDSS-IV acknowledges support and resources from the Center for High-Performance Computing at
the University of Utah. The SDSS web site is www.sdss.org.
\par
SDSS-IV is managed by the Astrophysical Research Consortium for the Participating Institutions of the SDSS Collaboration including the Brazilian Participation Group, the Carnegie Institution for Science, Carnegie Mellon University, the Chilean Participation Group, the French Participation Group, Harvard-Smithsonian Center for Astrophysics, Instituto de Astrof\'isica de Canarias, The Johns Hopkins University, Kavli Institute for the Physics and Mathematics of the Universe (IPMU) / University of Tokyo, Lawrence Berkeley National Laboratory, Leibniz Institut f\"ur Astrophysik Potsdam (AIP), Max-Planck-Institut f\"ur Astronomie (MPIA Heidelberg), Max-Planck-Institut f\"ur Astrophysik (MPA Garching), Max-Planck-Institut f\"ur Extraterrestrische Physik (MPE), National Astronomical Observatories of China, New Mexico State University, New York University, University of Notre Dame, Observat\'ario Nacional / MCTI, The Ohio State University, Pennsylvania State University, Shanghai Astronomical Observatory, United Kingdom Participation Group, Universidad Nacional Aut\'onoma de M\'exico, University of Arizona, University of Colorado Boulder, University of Oxford, University of Portsmouth, University of Utah, University of Virginia, University of Washington, University of Wisconsin, Vanderbilt University, and Yale University.

\bibliographystyle{mnras}
\bibliography{IMF} 

\bsp	
\label{lastpage}
\end{document}